\documentclass[preprint, a4paper,12pt]{article}

\usepackage{tikz}
\usepackage{amsmath,amssymb,graphicx,multirow,xspace,slashed}
\usepackage[linktocpage=true,colorlinks=true,urlcolor=blue,anchorcolor=blue,citecolor=red,filecolor=blue,linkcolor=blue,menucolor=black,pagecolor=blue]{hyperref}
\usepackage[compress,numbers]{natbib}
\usepackage{subfigure, placeins}
\usepackage{booktabs}
\usepackage{blindtext, rotating}
\usepackage{afterpage}
\usepackage{enumitem}
\usepackage{ marvosym }
\usepackage{authblk} 
\usepackage{verbatim}
\usepackage{soul} %scratch out text
\usepackage[normalem]{ulem}
\usepackage{pifont}% http://ctan.org/pkg/pifont
\usepackage[usestackEOL]{stackengine}
\usepackage{subfiles}
\newcommand\sq{\framebox(10,10){}\kern\fboxrule}

\setstackgap{S}{\fboxrule}

\usepackage[utf8]{inputenc}

\usepackage{listings}

\usepackage{floatrow}
\newfloatcommand{capbtabbox}{table}[][\FBwidth]

\usepackage[font=footnotesize,labelfont=bf]{caption}

\usepackage{lineno}

\allowdisplaybreaks

\long\def\symbolfootnote[#1]#2{\begingroup%
\def\thefootnote{\fnsymbol{footnote}}\footnote[#1]{#2}\endgroup}

\usepackage{hyperref}
\hypersetup{
    pdfnewwindow=true,      % links in new window
    colorlinks=true,       % false: boxed links; true: colored links
    linkcolor=blue,          % color of internal links
    citecolor=blue,        % color of links to bibliography
    filecolor=blue,      % color of file links
    urlcolor=blue        % color of external links
}

\newcommand{\newc}{\newcommand}
\newc{\gsim}{\lower.7ex\hbox{$\;\stackrel{\textstyle>}{\sim}\;$}}
\newc{\lsim}{\lower.7ex\hbox{$\;\stackrel{\textstyle<}{\sim}\;$}}
\newc{\gev}{\,{\rm GeV}}
\newc{\mev}{\,{\rm MeV}}
\newc{\ev}{\,{\rm eV}}
\newc{\kev}{\,{\rm keV}}
\newc{\tev}{\,{\rm TeV}}
\newc{\MHT}{$H_T^{\text{miss}}$}
\newc{\MET}{$\slashed{E}_T$}
\newc{\MTT}{$M_{T2}$}

\newc{\mz}{M_Z}
\newc{\mpl}{M_*}
\newc{\mw}{m_{\rm weak}}
\newc{\nr}[1]{N^c_R{}_{#1}}

\def\beq{\begin{equation}}
\def\eeq{\end{equation}}
\newcommand{\bea}{\begin{eqnarray}\begin{aligned}}
\newcommand{\eea}{\end{aligned}\end{eqnarray}}
\def\bitem{\begin{itemize}}
\def\eitem{\end{itemize}}

\definecolor{darkgreen}{rgb}{0,0.5,0}
\definecolor{goodyellow}{rgb}{0.9,0.7,0}

\numberwithin{equation}{section}

\newcommand\fverb{\setbox\fverbbox=\hbox\bgroup\verb}

\newbox\fverbbox

\begin{document}
\baselineskip 0.6cm

\begin{titlepage}
% what package is needed here?
%\preprint{LTH-1327}
\begin{flushright}
LTH-1327
\end{flushright}

\vspace{0.4 cm}
\thispagestyle{empty}

\begin{center}

\vskip 0.1cm
{\Huge \bf Scalar Co-SIMP Dark Matter:} 

\vskip 0.2cm

{\Huge \bf Models and Sensitivities}
\vskip 0.5cm

%\vskip 0.0cm
{\footnotesize
{ %\large John F. Beacom$^{1,2}$, 
\large Aditya Parikh$^{1,2}$, Juri Smirnov$^{3,4}$, W. Linda Xu$^{1,5,6}$, Bei Zhou$^{7}$}
\vskip 1.0cm
{\it $^1$ Department of Physics, Harvard University, Cambridge, MA 02138, USA \\}
{\it $^2$ C.N. Yang Institute for Theoretical Physics, Stony Brook University, NY 11794, USA \\}
{\it $^3$ Department of Mathematical Sciences, University of Liverpool,
Liverpool, L69 7ZL, UK\\}
{\it $^4$ The Oskar Klein Centre, Department of Physics, Stockholm University, AlbaNova, SE-10691 Stockholm, Sweden\\}
{\it $^5$ Berkeley Center for Theoretical Physics, Department of Physics, University of California, Berkeley, CA 94720, USA\\}
{\it $^6$ Theoretical Physics Group, Lawrence Berkeley National Laboratory, Berkeley, CA 94720, USA\\}
{\it $^7$ William H. Miller III Department of Physics and Astronomy, Johns Hopkins University, Baltimore, MD 21218, USA \\}
}

\vskip 0.3cm

\end{center}

\vskip 0.6cm

\begin{abstract}

In this work, we present UV completions of the recently proposed number-changing Co-SIMP freeze-out mechanism. 
In contrast to the standard cannibalistic-type dark matter picture that occurs entirely in the dark sector, the $3\to 2$ process setting the relic abundance in this case requires one Standard Model particle in the initial and final states. This prevents the dark sector from overheating and leads to rich experimental signatures. We generate the Co-SIMP interaction with a dark sector consisting of two scalars, with the mediator coupling to either nucleons or electrons. In either case, \textit{the dark matter candidate is naturally light}:  nucleophilic interactions favor the sub-GeV mass range and leptophilic interactions favor the sub-MeV mass range.  Viable thermal models in these lighter mass regimes are particularly intriguing to study at this time, as new developments in low-threshold detector technologies will begin probing this region of parameter space.
%they will imminently be probed by developing low-threshold technologies. 
%In this work we present possible UV-complete scenarios that have a dark matter candidate that freezes out by a recently proposed number-changing process, the Co-SIMP freezeout. 
%The key feature of this mechanism, in contrast to the standard cannibalistic picture, is that the process which sets the relic abundance couples to the Standard Model, thereby preventing the dark sector from overheating, and leads to observable signatures in laboratory experiments.
%Furthermore, the naturally expected dark matter mass range is low. 
%For nucleophilic interactions, the sub-GeV mass range is favored and for electrophilic scenarios the sub-MeV mass range is. 
While particles in the sub-MeV regime can potentially impact light element formation and CMB decoupling, we show that a late-time phase transition opens up large fractions of parameter space.  These thermal light dark matter models can instead be tested with dedicated experiments. We discuss the viable parameter space in each scenario in light of the current sensitivity of various experimental probes and projected future reach.

\end{abstract}

\flushbottom

\end{titlepage}

\setcounter{page}{1}

\tableofcontents

\vskip 1cm

\newpage

%%%%%%%%%%%%%%%%%%%%%%%%%%%%%%%%%%%%%%%%%%%%%
%%%%%%              SECTION            %%%%%%
%%%%%%%%%%%%%%%%%%%%%%%%%%%%%%%%%%%%%%%%%%%%%
\section{Introduction}
\label{sec:intro}

% low level 
The nature of dark matter (DM) is a longstanding open question in fundamental physics. Understanding its clustering properties will provide crucial information for astrophysics and cosmology, while analyzing its particle interactions will provide the next major clue regarding what lies beyond the Standard Model (SM) of particle physics. 

%From the theoretical point of view, the production mechanism in the early universe gives the strongest handle on a particular dark matter scenario. A major class of such production mechanisms is the thermal freezeout, its power lies in the agnosticism of the late time particle abundance to the exact conditions of the early universe at arbitrary large temperatures. 
%Nevertheless, at the temperature at which a particular interaction decouples the model has strong sensitivity to the conditions at that time. 

%  More detailed intro 
Among various particle physics models for dark matter, a powerful and predictive class is thermal-relic dark matter. Under the assumption that dark matter was in thermal contact with the hot SM plasma in the early universe, the strength of the interaction that keeps the thermal link between the dark and visible sectors will determine the dark matter abundance once this interaction freezes out. 
Since the total dark matter abundance is known today, the strength of the required interaction rate can be predicted given a specific number-changing mechanism, which is crucial for experimental validation. Furthermore, we expect that the detection of a thermally produced dark relic will provide a new window into the early universe, similar to the formation of light elements during Big-Bang Nucleosynthesis (BBN). 
%Intriguingly, recent studies have provided hints that light dark matter that was not in thermal contact with the SM plasma, would lead to isocurvature perturbations that are incompatible with the cosmic microwave background (CMB) observations~\cite{Bellomo:2022qbx}. 
%Thus, it is a crucial task to identify thermal alternatives that lead to sub-GeV dark matter candidates.

Various types of interactions that set the relic abundance are known. The most prominent, and the first to be proposed, is the WIMP freeze-out~\cite{Steigman:1984ac, Jungman:1995df, Bertone:2004pz, Bertone:2016nfn}. The relic abundance in this case is determined by the freeze-out of a $2 \rightarrow 2$ annihilation process of dark matter particles, and generically prefers a DM candidate with a mass around $100$ GeV. The predicted rate is a target for various indirect searches for dark matter annihilations in space, and WIMP model realizations often include some amount of DM-SM scattering that may be searched for with great sensitivities in controlled terrestrial experiments.  
Increasingly stringent experimental constraints in the WIMP parameter space, as well as the expansion of the direct and indirect detection programs to lower and higher mass regimes, has driven up the appetite for realizations of thermal relic dark matter in a wider mass range. In response, a number of alternate mechanisms have been studied. 
% Model/Scenario overview 
For example, in the case of dark matter co-annihilation with a color-charged partner, the thermal freeze-out can be achieved with very large dark matter masses~\cite{DeLuca:2018mzn,Gross:2018zha,Harz:2018csl} that can even exceed the unitarity bound~\cite{Griest:1989wd,Smirnov:2019ngs}. 
It has been shown in Refs.~\cite{Asadi:2021pwo,Asadi:2021yml} that thermal relics with masses well above the unitarity bound are expected in confining gauge theories that feature a first-order phase transition or significant entropy injection~\cite{Asadi:2021bxp,Asadi:2022vkc,Smirnov:2022tcg}. Furthermore, dark sector phase transitions have been shown to lead to compact dark matter candidates~\cite{Asadi:2022njl}, which could be the source for a newly identified class of supernova events~\cite{Smirnov:2022zip}. Finally, inelastic scattering against lighter particles in the thermal bath can also lead to new thermal relic targets~\cite{DAgnolo:2017dbv, DAgnolo:2019zkf}.

% Semi-annihilation
%Alternative reaction topologies have also been found to lead to a modified dark matter relic abundance.For example, semi-annihilation of dark matter particles and a dark sector mediator have been considered in Ref.~\cite{DEramo:2010keq}.
% SIMP

Other number-changing interactions have also been proposed: one of these is the SIMP scenario where the dark matter relic abundance is set by the freeze-out of an entirely dark interaction $\chi + \chi + \chi \rightarrow \chi  + \chi$. This opens a new avenue to realize thermally produced sub-GeV dark matter scenarios. Since the process that sets the relic abundance in this case resides entirely in the dark sector, it is relatively unconstrained and difficult to test experimentally~\cite{Hochberg:2014dra}. A number of further studies have considered variants of number-changing scenarios in the dark sector~\cite{Cline:2017tka, Kopp:2016yji, Farina:2016llk,Dey:2016qgf,DAgnolo:2018wcn, Kim:2019udq,Ghosh:2023ocl}, but the direct experimental validation of the interaction leading to freeze-out can be very challenging in these models. Furthermore, the excess buildup of thermal energy as a result of internal depletion within a monolithic dark sector can easily lead to hot dark matter and washout of structure formation. Thus, it is a crucial task to identify experimentally testable thermal alternatives that lead to sub-GeV dark matter candidates.

%however, overheating + non-predictability.

%  Co-SIMP
Most recently, a number-changing process that involves SM particles has been proposed in Ref.~\cite{Smirnov:2020zwf}. The intriguing features of this Co-SIMP framework are that the $\chi + \chi + \text{SM} \rightarrow \chi + \text{SM}$ reaction rate is directly predicted from the dark matter freeze-out condition, and that it can be tested in direct detection experiments. Furthermore, the Co-SIMP interaction kills two birds with one stone, by setting the relic abundance of dark sector particles to the value predicted by cosmology and by keeping the dark and visible sectors in kinetic equilibrium, thereby preventing the overheating of the dark sector through conversion of rest mass into kinetic energy.

\begin{figure}
    \caption{
   The Co-SIMP interaction topology~\cite{Smirnov:2020zwf}.
    }
    \resizebox{0.55\columnwidth}{!}{
    \includegraphics[scale=0.45]{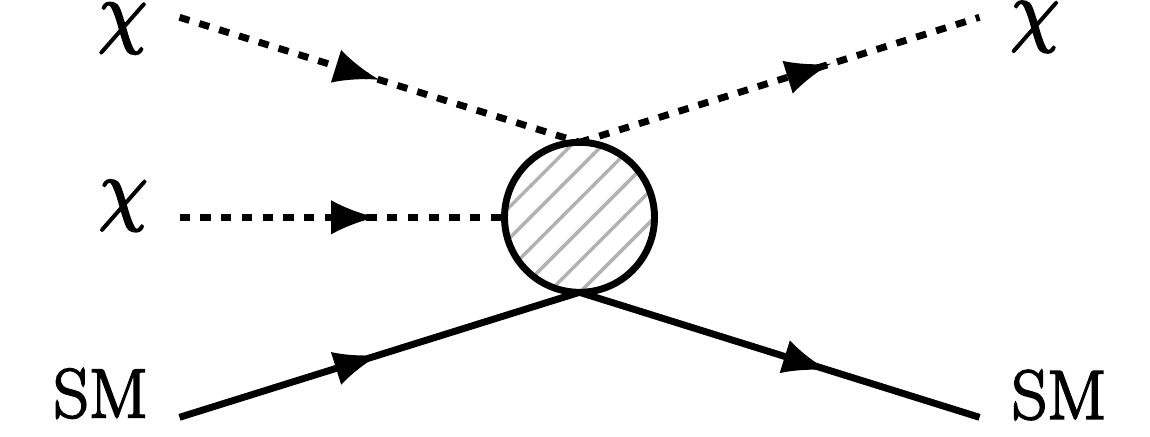}
       }
     \label{fig:cosimp}
\end{figure}

Fig.~\ref{fig:cosimp} shows the proposed reaction topology of the Co-SIMP interaction. In Ref.~\cite{Smirnov:2020zwf}, the Co-SIMP mechanism was proposed and investigated as an effective field theory (EFT). A coarse estimate of the relevant mass scales for the dark sector particles, assuming that the Co-SIMP interaction sets the relic abundance at freeze-out, gives us
\begin{align}
    \langle \sigma v^2 \rangle_{3\to 2} \equiv \frac{\alpha^2}{m_\chi^3 m_{\rm SM}^2}  \qquad m_\chi \sim \alpha^{2/3} (M_{ \rm pl} T_{\rm eq}^2)^{1/3} \sim \mathcal{O}(\mathrm{MeV})\,, \
\end{align}
where $T_{\rm eq}$ is the temperature at matter-radiation equality. Thus, the Co-SIMP framework provides scenarios of thermally produced sub-GeV dark matter candidates that directly involve SM particles in the production process. The freeze-out mechanism also points towards a target cross section range that can be probed in direct detection searches. This is a unique situation in dark matter physics and a particularly pertinent scenario to study now as experimental efforts are just beginning to explore these mass ranges (see Ref.~\cite{Essig:2022dfa} for a recent overview).
%\LX{Something to sell this as a thermal low-mass DM model probable via DiDt. }

However, since the EFT considered only the effective vertex, important cosmological effects and signatures at high-energy experiments due to the mediator were not thoroughly explored.
In this paper, we go beyond the EFT framework and suggest classes of models where the interaction in Fig.~\ref{fig:cosimp} is induced by a spin-0 bosonic mediator. As with any theory of light scalars, the models we consider come with their own naturalness and stability concerns.  Our models do not have a mechanism preventing the mediator $\phi$ from acquiring a vev, which in turn would induce a contribution to e.g. the electron mass of $\mathcal{O}(y_{\phi e} \langle \phi \rangle)$. However, our theory only relevantly couples to the light quarks or light leptons (in the nucelo- and lepto- philic cases respectively) in the SM, whose Higgs yukawas are not yet meaningfully measured.
%, known to a precision of $\mathcal{O}(10^{-8})$ keV, 
In a follow-up work~\cite{Co-SIMP_Vector}, we consider the case of vector mediators that are immune to many of the issues that make model-building with scalars difficult.

%which begs the concern for the stability and naturalness of these new light scalars. For instance, this model as stated does not provide protection against $\phi$ acquiring a vev, so some tuning is required in order to avoid the consequences of doing so. Most pertinently, such a vev would induce a contribution to the electron mass of order $y_{\phi e} \langle \phi \rangle$, which is known to $\mathcal{O}(10^{-8})$ keV precision. 
 
%With these caveats--broadly inherent to any theory with light scalars--in place, we can proceed to describe the cosmology and experimental constraints of this model. 

%Ultimately, the goal is to argue for a scenario that is viable albeit constrained, and furthermore to use this example to lay out and motivate the types of calculations and considerations we will use to assess other models, e.g. a vector-mediated fermion scenario.

%\LX{Other completions exist that potentially are not subject to these concerns, e.g. vector-mediation (which will be explored in future work). Sell the possibilities}\ap{added some words into a footnote}

The paper is organized as follows. In Secs.~\ref{sec:scalar_nucleo} and \ref{sec:scalar_lepto}, we discuss models where the dark sector couples to the SM via a scalar mediator. We demonstrate how, using late and inverse decays of the scalar mediator, the dark sector can come into equilibrium at the end of BBN and avoid severe constraints. This opens up a large fraction of light dark matter model parameter space. We discuss separately the coupling of the scalar mediator to nucleons (Sec.~\ref{sec:scalar_nucleo}) and leptons (Sec.~\ref{sec:scalar_lepto}).  Finally, we conclude in Sec.~\ref{sec:conclusions}.

%%%%%%%%%%%%%%%%%%%%%%%%%%%%%%%%%%%%%%%%%%%%%
%%%%%%              SECTION            %%%%%%
%%%%%%%%%%%%%%%%%%%%%%%%%%%%%%%%%%%%%%%%%%%%%
\section{Nucleophilic Co-SIMPs}
\label{sec:scalar_nucleo}
The simplest class of models that can realize the Co-SIMP freeze-out have a dark sector that consists of two particle species and is equipped with a $\mathbb{Z}_{3}$ symmetry that ensures DM stability.\footnote{We note that this symmetry can be elevated to a gauge symmetry. This is critical since quantum gravity does not admit global symmetries (see~\cite{Harlow:2022gzl} for a recent review) and generates Planck-suppressed operators violating them, which can in turn lead to DM decay. Instead, if this is a gauge symmetry, the higher dimensional operators respect the symmetry, rendering the DM absolutely stable~\cite{Baek:2013qwa, Ko:2014nha, Frigerio:2022kyu}.} The Co-SIMP dark matter is a complex scalar field $\chi$. The coupling to the SM is realized by a real scalar field $\phi$. In this section, we consider the scenario where all dark sector particles are scalar and the mediator couples the dark matter to quarks, or effectively SM nucleons.

\subsection{Model setup}

In order to generate a nucleophilic Co-SIMP scenario, we need a scalar field $\chi$ that for stability reasons has to enjoy a $\mathbb{Z}_{3}$ symmetry. The other ingredient is the mediator that couples this field to the SM particles. We choose a real scalar field $\phi$, which after electroweak symmetry breaking (EWSB) has a coupling to SM quarks $y_{\phi q} \phi \bar{q} q$. We first discuss how this coupling is generated. 

For the generation of the coupling, consider a new vector-like quark $F$, i.e. a particle where the left and right handed components, denoted $F_{L}$ and $F_{R}$ respectively, carry the same quantum numbers as the right-handed quark of the SM. Thus, an explicit mass term is allowed $M_F \bar{F} F$ and no contribution to the triangle anomalies is generated. Furthermore, the theory contains the following relevant terms: the coupling to the mediator $y_{\phi}^{q} \phi \bar{F}_{L} q_R$, where $q_R$ is the right-handed SM quark and the coupling to the Higgs $y_{H}^{q} H \bar{Q}_L F_{R}$, where $Q_L$ is the left-handed SM quark doublet. Finally, $\tilde{M} \bar{F}_{L} q_R$, another explicit mass term, is also permitted.  These four terms, along with the dark sector interaction $\phi \chi^3$,  fully describe the relevant interactions of our theory at high energies.  $\phi$ and $\chi$ are not charged under the SM gauge group, and though $F$ is, we assume its mass is well above the EW scale $M_F \gg v_H$ so it does not directly contribute to the phenomenology. Other than the DM itself, none of the new particles are charged under the dark discrete symmetry.

Thus we have after EWSB an induced mediator coupling to SM quarks $\bar{q}q$ given by
\begin{align}
    \frac{y_{H}^q y_{\phi}^q v_H }{M_F} \, \phi \bar{q} q = y_{\phi q} \phi \bar{q} q \,. 
\end{align}
In addition, this construction induces a quark mass shift of order $\Delta m_q \sim y_{H}^q v_H \tilde{M}/M_F$, but with the requirement that $\tilde{M} < \text{GeV}$ this effect remains negligible. 
%\js{add cite for sensitivity? Future search?} \LX{personally inclined that more elaboration/citations here isn't necessary -- this is pretty far out of reach for even b/t yukawa measurements, and we basically later assume only coupling to light quarks. }

\begin{figure}[h!]
    \caption{
   DM depletion mechanisms in the nucleophilic Co-SIMP case.  We keep the third process, which is more important for the leptophilic scenario (Sec.~\ref{sec:scalar_lepto}), here for comprehensiveness.
    }
    \resizebox{\columnwidth}{!}{
    \includegraphics[scale=0.45]{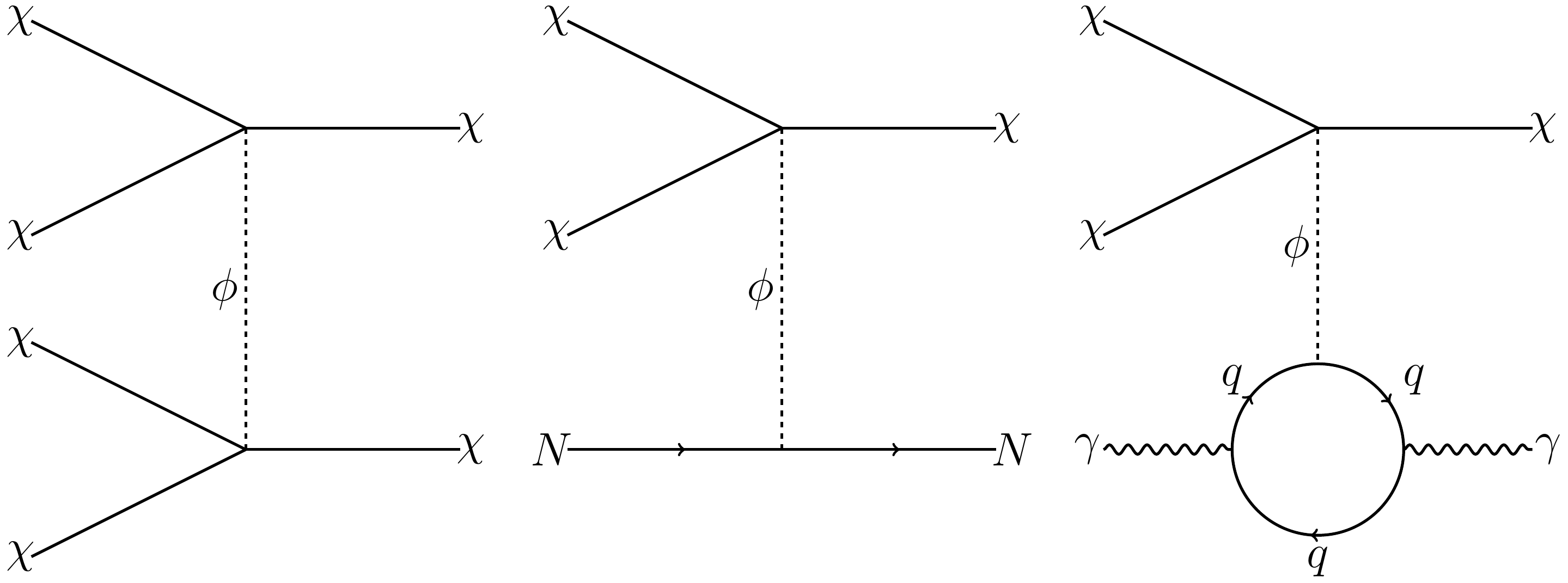}
       }
     \label{fig:dm_depletion_nucl}
\end{figure}

Thus, after EWSB our model is described by the following Lagrangian,
\begin{align}
\mathcal{L} &\supset  \partial_\mu \chi^\dagger \partial^\mu \chi +  \frac{1}{2} \left(\partial_\mu \phi\right)^2  -  m_\chi^2 |\chi|^2  -  \frac{1}{2} m_\phi^2 \phi^2 + \sum_q y_{\phi q} \phi \bar{q} q + \frac{y_{\phi \chi}}{3!} \phi \chi^3 - V(\chi, \phi)\,,
\label{eqn:scalar_lagrangian_nucl}
\end{align}
 where the potential can contain all other generically allowed renormalizable scalar operators
\begin{equation}
\begin{split}
V(\chi,\phi) &\supset \frac{\lambda_1}{4} |\chi|^4 +  \frac{\lambda_2}{2} |\chi|^2 \phi^2  + \frac{\lambda_3}{4!} \phi^4  + \lambda_4 |H|^{2} |\chi|^2 + \frac{\lambda_5}{2} |H|^2 \phi^2 \\
&+\frac{\mu_1}{3!} \chi^3 + \mu_2 |\chi|^2 \phi + \frac{\mu_3}{3!} \phi^3 + \mu_{4}|H|^{2}\phi + \text{\rm h.c.},
\label{eqn:scalar_potential}
\end{split}
\end{equation}
The stability of the dark matter (but not the mediator) is ensured by a $\mathbb{Z}_{3}$ charge. %$\phi$ is neutral and $\chi$ has charge 1 under the $\mathbb{Z}_{3}$. 
In this scenario, the topology of the Co-SIMP interaction is realized in a $\chi\chi N \to \chi^\dagger N$ process below the QCD scale, with the caveat that the outgoing DM particle is of opposite $\mathbb{Z}_{3}$ charge as the incoming two. The corresponding diagram is shown in Fig.~\ref{fig:dm_depletion_nucl}, along with a one-loop counterpart where the interaction occurs with photons.

The couplings relevant to the Co-SIMP process specifically are $y_{\phi\chi}$ and $y_{\phi q}$, but other terms included in $V(\chi, \phi)$ are generically allowed to appear and may influence the overall phenomenology. In particular, it is worth noting that based on the charge assignments for the dark sector fields, additional Higgs portal interactions exist which can impact freezeout as well as the direct detection rate. Furthermore, there is a rich vacuum structure associated with the potential shown. Different choices of the trilinear and quartic couplings may lead to the existence of false vacua which can have interesting phenomenological consequences of their own. We make the notational delineation between the ``Co-SIMP parameters" $y$ and the ``free parameters" $\lambda$ and $\mu$ to emphasize that only the former need to be sufficiently large to ensure a Co-SIMP dominated cosmology that freezes out to the right relic abundance. In addition, we tune the values of the remaining parameters such that constraints can be avoided as necessary. It should be noted here that our observables are actually insensitive to $\phi$ acquiring a vev,   since the only true consequences would be a shift to the light quark masses (and thus the implied Higgs yukawas, which are not explicitly measured).

Consequently, in the following discussion we neglect the terms included in the self-interaction potential $V(\chi,\phi)$, assuming they are sufficiently small to avoid undesirable phenomenology. The requirements on the sizes of the ``Co-SIMP parameters" set by the present-day DM abundance are discussed in the following subsection. 

To find the low-energy effective description of the dark sector interacting  with nucleons, we follow the formalism in Ref.~\cite{Cline:2013gha} and define the effective coupling strength by
\begin{align}
    \mathcal{L}_{\rm int}^{\rm eff} =  y_{N} \phi \bar{N} N \qquad      y_{N} \equiv \sum_q y_{\phi q} f_q = \sum_q y_{\phi q} \langle N| \bar{q} q  |N \rangle \,.
\end{align}

As we discuss below, we assume that the mediator dominantly couples to light quarks, so we are particularly interested in determining those matrix elements. The matrix elements for the light quarks can be expressed as 
\begin{align}
& f_u = \frac{\sigma_l}{m_u + m_d} \frac{2 z+ \zeta (1-z)}{1+z} \approx 9.2\\ \nonumber
& f_d = \frac{\sigma_l}{m_u + m_d} \frac{2 - \zeta (1-z)}{1+z} \approx 6.3\\ \nonumber 
& f_s = \frac{\sigma_l}{m_u + m_d}\, \zeta \approx 0.4\,.
\end{align}
Here, $z$ quantifies the isospin breaking and was estimated in Ref.~\cite{Cline:2013gha} to be $z \approx 1.49$. 
The strangeness content parameter is defined as
\begin{align}
\zeta \equiv    \frac{\langle N| \bar{s} s  |N \rangle }{\langle N| \bar{u} u +  \bar{d} d |N \rangle } = 1 - \frac{\sigma_0}{\sigma_l}\,.
\end{align}

The relevant parameters $\sigma_0$ and $\sigma_l$, needed for the numerical evaluation above, are determined from measurements and lattice simulations.  These values are given by  $\sigma_0 = 55 \pm 9 \, \rm MeV $ and $\sigma_l = 58 \pm 9 \, \rm MeV$ in Ref.~\cite{Cline:2013gha}. 

In order to evaluate the heavy quark contributions to the nucleon coupling, we follow the strategy suggested in Ref.~\cite{Shifman:1978zn}. First, by matching the trace anomaly of the energy-momentum tensor at high and low energies, we find that 
\begin{align}
1 - f_N =   - \frac{7}{8 \pi\, m_N} \langle N| G_{\mu \nu} G^{\mu \nu} |N \rangle \, ,
\end{align}
where $f_N \approx 0.3$ is the Higgs-nucleon coupling. Second, for the heavy quarks, the matrix element with nucleons is generated by the gluonic operators and is given by
\begin{align}
\langle N| \bar{q} q  |N \rangle =   - \frac{1}{12 \pi\, m_q} \langle N| G_{\mu \nu} G^{\mu \nu} |N \rangle \, ,
\end{align}
for the quark flavors $q  = c, b,t$. Combining those equations allows us to express the heavy quark nucleon coupling constants through the Higgs-nucleon coupling
\begin{align}
f_q =  \frac{2 \, m_N}{21 \, m_q} \left( 1 - f_N \right) \, ,
\end{align}
which can be evaluated to obtain
\begin{align}
f_c =  0.049, \, f_b =  0.015, \, f_t =  3.6\times 10^{-4}\, .
\end{align}
Thus, we note that the Co-SIMP-nucleon coupling is driven by the interaction with the up and down quarks. For example, assuming a model with $y_{\phi u} = y_{\phi d} = \epsilon $ and $y_{\phi i} = 0$ for the rest, we have $y_{N} \approx 16 \, \epsilon$. We call this model, \textbf{Model I}, and use it in the following sections for demonstration purposes, with $y_{N} $ related to $y_{\phi u, \phi d}$ as discussed above.

With the connection between the fundamental parameters of the theory and effective nucleon couplings set, the sensitivity of different experiments to the parameter space can be explored. 
%Note that since the scalar field $\phi$ does not develop a vev, there is no contribution to the nucleon mass and trace anomaly. 
Finally, we can ignore the coupling to pions as the derivative interactions lead to velocity suppression, rendering the pion channels subdominant.

%%%%%%%%%%%%%%%%%%%%%%%%%%%%%%%%%%%%%%%%%%%%%%%%%%%%%%%%%%%%%
%%%%%%%%%%%%%%%%%%%%%%%%%%%%%%%%%%%%%%%%%%%%%%%%%%%%%%%%%%%%%
\subsection{Early Universe Cosmology}

%%%%%%%%%%%%%%%%%%%%%%%%%%%%%%%%%%%%%%%%%%%%%%%%%%%%%%%%%%%%%

In this scenario, we require the dark sector to efficiently interact with SM nucleons; as such, the typical mass ranges for these particles are at the sub--GeV scale. 
%The leptophilic scenarios described in later sections are concerned with much lighter DM masses and are subject to additional constraints such as those from BBN and stellar cooling, and we refer the reader to discussions therein for those details. 
The strength of the dark sector couplings is primarily determined by the requirement of thermal freeze-out, assuming Co-SIMPs constitute 100\% of the observed dark matter abundance.  We use the freeze-out constraint to identify the required model parameters and compare them to current and future experimental sensitivities in the coming sections. 

Two broad scenarios need to be considered separately. On the one hand, the dark sector can be in thermal equilibrium at high energy scales and the freeze-out of the Co-SIMP process sets the dark matter relic abundance. On the other hand, thermalisation can be delayed to much later times and lower energies. Even though the Co-SIMP process sets the dark matter abundance in this case as well, the impact on early universe observables is drastically different. We discuss the early thermalization scenario in the coming section, and then consider late thermalization.

%The Boltzmann equations governing the number-changing processes are given by 
%\begin{align}
%    \frac{d Y_\chi}{ d x} = \frac{s}{x H}
%\end{align}
%where $Y \equiv n/s,  x \equiv m/T$ in the forward-dominated regime $n_\chi \gg n_{\chi, \rm eq}$ 

%%%%%%%%%%%%%%%%%%%%%%%%%%%%%%%%%%%%%%%%%%%%%%%%%%%%%%%%%%%%%
\subsubsection{Early Thermalization, Freeze-out and Relic Abundance}

A cornerstone property of thermal dark matter is that the present-day dark matter abundance is set by the freeze-in or freeze-out of number-changing interactions in the early universe. For the purpose of this article, we are primarily interested in scenarios where the abundance is set specifically by the freeze-out of Co-SIMP processes, but for any given UV completion such as this one, it is often the case that several different processes are available to contribute to DM depletion. To a good approximation, the epoch of freeze-out is simply set by the most efficient one.  Regardless, it is potentially the case that different regions of parameter space for a given model correspond to different processes dominantly setting the relic abundance. In this subsection, we attempt to investigate the various processes that could ostensibly dominate the freeze-out process, and build intuition for the regions of parameter space where the Co-SIMP process is the dominant one. We assume that the reheating temperature is large enough to efficiently produce dark sector particles and that interactions between the dark and SM sectors thermalize them, such that there is only one common temperature $T$.

While the Co-SIMP process is kinematically forward-dominated and naturally always contributes to the depletion of dark matter, it may be out-competed by other diagrams with larger couplings or weaker number density dependencies. To focus on the parameter space where the relic abundance is driven by this $3\to 2$ topology, we restrict our consideration to regions where $m_\chi < m_\phi$ (for our analysis we assume that $m_\chi < 0.75 \, m_\phi$ such that even at finite temperature, mediator emission is suppressed) and $m_\chi \lesssim m_\pi$, such that semi-annihilation $\chi\chi \to \chi^\dagger \phi$ and 2- (and 3-) annihilation $\chi \chi^\dagger (\chi\chi\chi) \to \phi^* \to \pi \pi $ are kinematically disallowed, respectively. Note that the second requirement can likely be relaxed to $m_\chi \lesssim \text{GeV}$, as the pion coupling is suppressed and it suffices to  kinematically forbid $\chi \chi^\dagger (\chi\chi\chi) \to \phi^* \to  N N$, with $N$ being a proton or neutron.
The annihilation(s) $\chi \chi^\dagger (\chi\chi\chi) \to \gamma\gamma$ are always allowed at one-loop, but are expected to be quite suppressed. The mediator $\phi$ is unstable since it can always decay to $\gamma\gamma$ at one-loop, but may have additional decay channels available depending on how heavy it is. We reiterate here that we assume the only couplings that appreciably contribute are $\{y_{N}, y_{\phi\chi} \}$, as these are the only terms that {\it need} to be large to support Co-SIMP annihilation; consequently, all diagrams including other dark sector couplings may be neglected. 

At last, the diagrams we consider as potentially important in setting the relic abundance are shown in Fig.~\ref{fig:dm_depletion_nucl} -- in addition to the nucleon-mediated tree-level Co-SIMP process, we consider the similar photon-mediated diagram at one-loop, whose loop and $\alpha^2$ suppression is potentially compensated by the large abundance of photons relative to nucleons in the universe.  Also potentially relevant is the DM-only $4\to2$ SIMP process. Although this process is controlled only by the dark sector couplings, which are generically much larger, it suffers severe phase space suppression and contributes to an overheated dark sector. 

While the dark sector is thermalized with the photon bath, the Boltzmann equations governing the abundances are given by

% \begin{figure}[h!]
%     \caption{
%   Possible DM depletion mechanisms.
%     }
%     \resizebox{0.8\columnwidth}{!}{
%     \includegraphics[scale=0.45]{figures/scalar_DMDMDMDM_to_DMDM.pdf}\hspace{4em}
%     \includegraphics[scale=0.45]{figures/scalar_DMDMelec_to_DMelec.pdf}\hspace{4em}
%     \includegraphics[scale=0.45]{figures/scalar_DMDMgamma_to_DMgamma.pdf}
%       }
%      \label{fig:dm_depletion}
% \end{figure}

%In Fig.~\ref{fig:dm_depletion_nucl} the dominant diagrams for Co-SIMP depletion are shown. 

%In this reduced parameter space, the leading diagrams that might dominate the epoch of freeze-out are: semi-annihilation $\chi\chi \to \chi^\dagger \phi$, Co-SIMP annihilations at tree-level and one-loop $\chi \chi N \to \chi^\dagger N$ and $\chi \chi \gamma \to \chi^\dagger \gamma$, 3-annihilations $\chi \chi \chi \to \bar{N} N$ and $\chi \chi \chi \to \gamma \gamma$, the DM-only SIMP annihilation $\chi \chi^\dagger \chi \chi^\dagger \to \chi \chi^\dagger$, and, if kinematically allowed, a loop-induced annihilation $\chi \chi^\dagger \rightarrow N \bar{N}$. The semi-annihilation process, if allowed to occur, generically dominates the DM depletion due to the relatively unconstrained nature of dark-sector-only couplings, so we kinematically forbid it by imposing $m_\chi < m_\phi$. We find in practice that the 3-annihilations contribute negligible support for thermal relics, and neglect them here. Furthermore, the loop induced annihilation process is suppressed in \textbf{Model I}, due to the light quark masses that dominate the coupling.  

%
\begin{align} 
\label{eq:boltzmann}
\frac{dn_\chi}{dt} + 3Hn_\chi  = & 
- \gamma_{\chi\chi N \to \chi N} \left[ \left(\frac{n_\chi}{n_{\chi, \rm eq}}\right)^2 \left(\frac{n_N}{n_{N, \rm eq}}\right) - \left(\frac{n_\chi}{n_{\chi, \rm eq}}\right) \left(\frac{n_N}{n_{N, \rm eq}}\right) \right] \nonumber\\
& - \gamma_{\chi\chi \gamma \to \chi \gamma} \left[\left(\frac{n_\chi}{n_{\chi, \rm \rm eq}}\right)^2 \left(\frac{n_\gamma}{n_{\gamma, \rm eq}}\right) - \left(\frac{n_\chi}{n_{\chi, \rm eq}}\right) \left(\frac{n_\gamma}{n_{\gamma, \rm eq}}\right) \right] \\
&- 2 \gamma_{4\chi \to 2\chi} \left[\left(\frac{n_\chi}{n_{\chi, \rm eq}}\right)^4 - \left(\frac{n_\chi}{n_{\chi, \rm eq}}\right)^2 \right] \, , \nonumber\\
\frac{dn_\phi}{dt} + 3Hn_\phi  = &  - \gamma_{\phi \to \gamma\gamma} \left[ \left(\frac{n_\phi}{n_{\phi, \rm eq}}\right) - 1 \right] \,.  
\end{align}
The effective interaction density is given by e.g. $\gamma_{\chi \chi N \to \chi N} \equiv \langle \sigma v^2 \rangle n^2_{\chi, {\rm eq}} n_{N, {\rm eq}} $. Just before the epoch of freeze-out as the DM turns non-relativisitic, the depletion process dominates, $n_\chi \gg n_{\chi, \rm eq}$, and we can further write
\begin{align} 
\frac{dn_\chi}{dt} + 3Hn_\chi & =  - \langle \sigma v^2 \rangle_{\chi\chi N \to \chi N}  n_\chi^2 n_N - \langle \sigma v^2 \rangle_{\chi\chi \gamma \to \chi \gamma} n_\chi^2 n_\gamma - 2 \langle \sigma v^3 \rangle_{4\chi \to 2\chi} n_\chi^4 \, .
\end{align}
The cross-sections for these processes are given, in the low-temperature limit, by
\begin{align}
\langle \sigma v^2 \rangle_{\chi\chi N \to \chi N} 
& = \frac{ \sqrt{3} y_{\phi\chi}^2 y_{N}^2 m_\chi^{-1}}{ 2 \pi m_\phi^2 (m_\phi^2 + \Gamma_\phi^2)  }  + \mathcal{O} \left( \frac{m_\chi}{m_N } \right) \, ,\\
\langle \sigma v^2 \rangle_{\chi\chi\gamma \to \chi \gamma}  
& =\sum_q \frac{\pi  y_{\phi q}^2 y_{\phi \chi}^2 \alpha^2 T}{30 \sqrt{3} \zeta(3) m_q^2 m_\phi^2 (m_\phi^2 + \Gamma_\phi^2) }      + \mathcal{O}(T^2) \, , \\
\langle \sigma v^3 \rangle_{4\chi\to 2\chi} \label{eq:crosssectionsNucleo}
& =  \frac{\sqrt{3} y_{\phi\chi}^4}{256 \pi m_\chi^4 \left[ (3 m_\chi^2 + m_\phi^2)^2 + m_\phi^2 \Gamma_\phi^2 \right] } \, ,\\ 
\Gamma_\phi & = \frac{ y_{N}^2 \alpha^2 m_\phi^3}{144 \pi^3 m_N^2 } + \mathcal{O}\left(\frac{m_\phi^4}{m_N^3}\right) . \label{eqn:phidecay}
\end{align}
We find that while for sufficiently small DM-SM couplings, the DM-only $4 \to 2 $ SIMP process dominates, this mode of depletion is not efficient enough in the mass range of interest to achieve the correct relic abundance for couplings that respect perturbative unitarity. Thus for the parameter space delineated above, we may solve the above system of equations to obtain a relic abundance dominated by Co-SIMP depletion. This relic abundance is given to a good approximation by 

\begin{align}
    \left( \frac{\Omega_\chi h^2}{0.12} \right) & \simeq \left( \frac{m_\chi}{\rm MeV}\right)^{-3} \left[ \frac{\eta \langle \sigma v^2 \rangle_{\chi \chi N \to \chi N}  + \langle \sigma v^2 \rangle_{\chi\chi \gamma \to \chi \gamma}}{ 0.01 \, {\rm MeV}^{-5} } \right]^{-1} \nonumber\\ 
    & \simeq  \left[ \frac{y_N}{0.1}\right]^{-2} \left[ \frac{y_{\phi\chi}}{0.1}\right]^{-2} \left( \frac{m_\chi}{\rm MeV}\right)^{-3}   \left(\frac{m_\phi}{100 {\rm MeV}} \right)^{4}  \nonumber \\
    &  \qquad \times  \left[ 10^{-5} \left( \frac{m_\chi}{\rm MeV}\right)^{-1} + \left(\frac{m_\chi}{\rm MeV} \right)  \right]^{-1}, 
\end{align}
where $\eta$ denotes the baryon-to-photon ratio, and we have, as promised, used the {\bf Model I} configuration of quark-level couplings. However, we note that this dependence is merely demonstrative and any specific choice of nucleon couplings may be very easily mapped to our results.

%%%%%%%%%%%%%%%%%%%%%%%%%%%%%%%%%%%%%%%%%%%%%%%%%%%%%%%%%%%%%
%%%%%%%%%%%%%%%%%%%%%%%%%%%%%%%%%%%%%%%%%%%%%%%%%%%%%%%%%%%%%

%\subsection{Early Universe Cosmology}
\subsubsection{Late Thermalization, Freeze-out and Relic Abundance}

In this subsection, we discuss the cosmology of a particle model that thermalizes late with the SM and then freezes out. As in the previous section, we assume throughout that the internal thermalization of the dark sector is efficient, and thus the entire dark sector (de-)couples with the SM all at once. In this scenario, the dark sector does not violate CP and the relic population is the explicit result of thermal production by SM particles, so the species is necessarily symmetric. 

As such, many pieces of this story, including the dynamics at freeze-out and the relationship to present-day abundance, are similar to the previous, thus we focus on the additional considerations brought about by late thermalization. The scenario relies on the following ingredients, demanding that the dark sector is:

\begin{itemize}
    \item not populated at very early times which can be most easily achieved by suppressing its coupling to the inflaton.
    \item subject to a phase transition  which affects the masses of the mediator, and the dark sector particles, and makes it light only at late times.
\end{itemize}

\begin{figure}
    \caption{
  Example potentials for the heavy dark scalar $\Phi$, that would lead to a phase transition with a low lying vacuum expectation value at late times. For a phase transition at $T < \rm MeV$, the energy release, which is controlled by $\epsilon$, does not significantly affect the dynamics of the early universe. 
    }
    \resizebox{0.6\columnwidth}{!}{
    \includegraphics[scale=0.45]{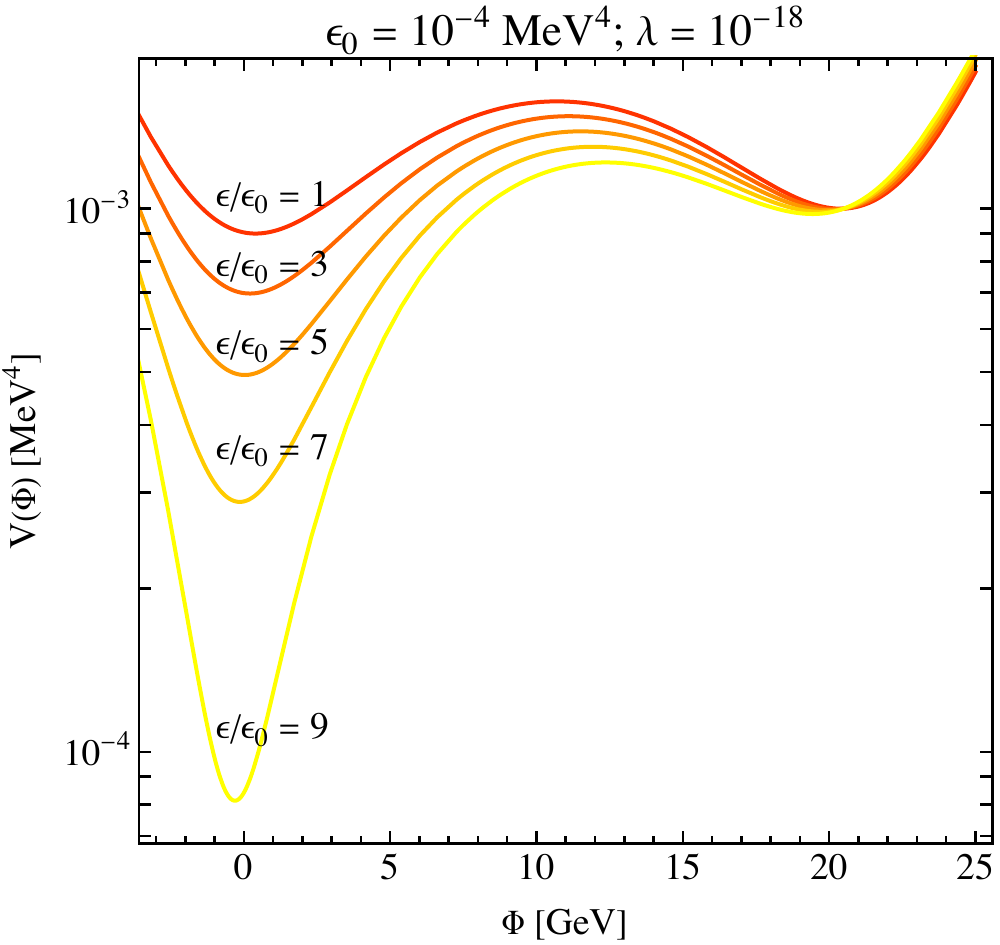}
       }
    \label{fig:potential}
\end{figure}

This is realized by the following mechanism. First, we introduce a second scalar $\Phi$ which couples to both the scalar mediator $\phi$ via $\lambda_\phi \, \phi^2\,  \Phi^2$ and the DM via $\lambda_\chi |\chi|^2 \Phi^2$. $\Phi$ subsequently gets a vacuum expectation value which fixes $m_\phi \approx \sqrt{\lambda_\phi} \langle \Phi \rangle$ and $m_\chi \approx \sqrt{\lambda_\chi} \langle \Phi \rangle$.
As an explicit example, we consider the following potential for $\Phi$ (illustrated in Fig.~\ref{fig:potential})
\begin{align}
    V(\Phi)  = \frac{\lambda}{8} \left[ M_\phi +\delta  -\Phi  \right]^2 \left( \Phi - \delta \right )^2 - \frac{\epsilon}{M_\phi}\left[ M_\phi +\delta  -\Phi  \right] .
\end{align}

Assuming that $\delta/M_\phi \ll 1 $ and $\epsilon/M_\phi^4 \ll 1$ are small parameters, the two minima of the potential are at $\langle \Phi \rangle_{\rm low} \approx \delta $ and $\langle \Phi \rangle_{\rm high} \approx M_\phi$. The potential difference between the minima is $\Delta V\approx \epsilon$. Note that dark sectors with increasing interaction strength, after a late time phase transition were considered in Refs.~\cite{Elor:2021swj,Mandal:2022yym}. 

We proceed with the analysis of the phase transition, based on the seminal works~\cite{Coleman:1977py,Callan:1977pt,Coleman:1980aw}, using the semi-classical approximation. The space-time phase conversion rate is given by

\begin{align}
    \gamma \approx \frac{B^2}{4 \pi R_c^4} \exp{\left[-B\right] } \qquad  B = \frac{27 \pi^2 S_1^4}{2 \epsilon^3}\,.
\end{align}
Here, the scale is set by the size of the bubble with a critical radius, i.e. the bubble radius where the volume energy wins over the surface tension, which leads to rapid expansion. This size is $R_c = 3 S_1/\epsilon$, with $S_1 \approx \sqrt{\lambda} \, M_\phi^3/24$. 

The time at which the phase transition largely completes and the true vacuum dominates the universe is $ \tau_{\rm PT}  \approx H(\tau_{\rm PT}  )^3/\gamma$, with $H(\tau_{\rm PT}  ) \approx T_{\rm PT}^2/M_{\rm pl}$, the Hubble rate at the time of the phase transition, assuming that the universe is radiation dominated. 

As discussed in Ref.~\cite{Coleman:1977py}, the energy gained in the conversion between phases is predominantly stored in the wall energy. This energy is then converted to gravitational radiation when the bubble walls collide. In the case that $\epsilon < T_{\rm SM}^4$, the universe at all times is dominated by the SM,  and the amount of gravitational radiation contributes negligibly to the expansion of the universe such that $\Delta N_{\rm eff} \approx 0$. 

Given that the dark sector has negligible coupling to the inflaton, and the masses of dark sector particles before the phase transition are above the reheating temperature, the dark sector is not populated before the phase transition. Even if operators of the type $\lambda_{\chi H} |\chi|^2 |H|^{2}$ are present, no freeze-in type population~\cite{Hall:2009bx} takes place. Note furthermore, that since the dark sector is not reheated after inflation, thermal corrections to the potential are not relevant prior to the phase transition. 

This mechanism allows for the realization of a wide variety of thermal light dark sectors that nonetheless conform with BBN constraints in the very early universe, including the leptophilic scalar Co-SIMP. Possibilities such as these illustrate the necessity of complementary dark sector probes from cosmology, astrophysics, and terrestrial detection -- the latter two of which are discussed in the following subsections. Along this vein, it is interesting to note that upcoming GW experiments can potentially probe the stochastic GWs produced during this type of phase transition~\cite{Sachdev:2020bkk, Sharma:2020btq, Biscoveanu:2020gds, Zhou:2022otw, Zhou:2022nmt, Zhong:2022ylh, Racco:2022bwj, Pan:2023naq}.

%%%%%%%%%%%%%%%%%%%%%%%%%%%%%%%%%%%%%%%%%%%%%%%%%%%%%%%%%%%%%
\subsubsection{Big-Bang Nucleosynthesis and the CMB}
%

%%%%%%%%%%%%%%%%%%%%%%%%%%%%%%%%%%%%%%%%%%%%%%%%%%%%%%%%%%%%%

An intriguing feature of new, thermally produced, sub MeV-scale particles is that they persist as radiation in the early universe until after the freeze-out of neutrons and decoupling of neutrinos from the SM plasma. Our current understanding of processes in the early universe allows us to restrict certain scenarios with these relativistic degrees of freedom. 

An important ingredient is the production of light nuclei during big-bang nucleosynthesis (BBN). Our precise understanding of the physics during the BBN epoch provides us with strong restrictions on allowed scenarios since the deuterium and Helium-4 abundances are affected by the number of light degrees of freedom.  A convenient parametrization of relativistic degrees of freedom is the number of effective neutrino species, which is constrained by BBN to be $N_{\rm eff} = 2.85 \pm 0.28 $ ($1\sigma$)~\cite{Hsyu:2020uqb}; this disfavors new, thermally populated species during this epoch with masses below $ \sim 0.7 \,  T_{\rm BBN}$ at the 2$\sigma$ level.  However, this limit does not apply if thermal production of dark sector particles from the SM bath is inefficient until well after $T_{\rm BBN} \approx 1 \text{ MeV}$~\cite{Berlin:2019pbq}; in other words, the dark sector enters thermal equilibrium with the SM after the epoch of neutrino decoupling.

Another observational anchor on the radiation content of the universe comes from anisotropies of the CMB. However, since the epoch of recombination occurs significantly later in the universe, at temperatures far below the Co-SIMP mass (and thus the temperatures of freeze-out), the impact of light Co-SIMPs on these observables is more indirect.  If the Co-SIMP sector thermalizes with the SM after the neutrinos have decoupled, the number of relativistic degrees of freedom is only affected if the thermalization takes place before the electrons fully freeze-out. That is indeed the case for leptophilic Co-SIMPs with masses slightly below the electron mass.  Nonetheless, because the entropy of the dark sector is largely returned to the SM plasma at freeze-out, the impact on $N_{\rm eff}^{\rm CMB}$ is negligible.
%, leading to $\Delta N_{\rm eff}^{\rm CMB} \sim 0$ for species that thermalize before and freeze-out after $e^+ e^-$ annihilation. 
However, if the Co-SIMP sector is populated prior to thermalization, or the entropy during Co-SIMP annihilation is partially transferred to the neutrino sector, next generation CMB experiments~\cite{CMB-S4:2016ple} could be sensitive to such  scenarios. 

%%%%%%%%%

BBN harbors great potential for testing models of light DM~\cite{Depta:2019lbe,Sabti:2019mhn}. Two observables of particular importance are the proton yield, $Y_P$, and the deuteron fraction, $D/H_P$. Thermalized, light species have two ways in which they affect these observables in general. On one hand, the light thermal degrees of freedom behave as radiation, affecting the Hubble rate, which in turn shortens the timescales on which weak and nuclear processes freeze-out, leading to larger $Y_{P}$ and $D/H_{P}$.
On the other hand, the typical freeze-out process $\chi + \chi \rightarrow \text{SM} + \text{SM}$ injects entropy into the SM which dilutes the baryon abundance and reduces $Y_P$ and $D/H_P$. 

In the Co-SIMP scenario, the entropy injection by the process $ \text{SM} + \chi + \chi  \rightarrow \,  \text{SM} + \chi$ is severely reduced as the SM particles mostly play a catalyzing role. Therefore, the main sensitivity to a sub-GeV Co-SIMP arises from its effect on the Hubble expansion rate during BBN. As discussed in Ref.~\cite{Sabti:2019mhn}, this excludes a fully thermalized particle with electromagnetic interactions below the mass of $10 \rm \, MeV$.  We note, however, that this constraint might be relaxed or evaded entirely by introducing non-minimal cosmological histories, which are explored in more detail in the leptophilic scenario.

%%%%%%%%%%%%%%%%%%%%%%%%%%%%%%%%%%%%%%%%%%%%%%%%%%%%%%%%%%%%%
\subsection{Astrophysical Reach}
\label{sec:scalar_leptophilic_astrophysical_constraints}
The major astrophysical constraints which could affect the parameter space of our model are dark matter self-scattering and stellar cooling.  Since our mediator and DM candidate are both roughly in the MeV--GeV range, stellar interiors, which are typically at temperatures of $\mathcal{O}$(100 keV) or lower, are not energetic enough to support a large thermal population of our dark sector particles~\cite{Raffelt:1996wa}. Thus, bounds from stellar cooling are avoided in the bulk of our parameter space. In the edge case where the mediator is very light, we consider bounds from red giant star cooling~\cite{Grifols:1986fc,1989MPLA....4..311G}. In addition to stellar cooling constraints, anomalous energy loss during a supernova explosion could have sensitivity to new light particles~\cite{DeRocco:2019jti}. However, as discussed in Ref.~\cite{Smirnov:2020zwf}, the $\chi + N \rightarrow \chi + \chi + N$ interactions lead to such a short mean free path inside the proto-neutron star that Co-SIMPs are effectively trapped and no significant amount of energy is lost. 
 
Another avenue for inferring the existence of new dark sector particles, absent of non-gravitational interactions with the SM, is via observations of dark matter self-scattering. The various interactions in Eq.~\ref{eqn:scalar_lagrangian_nucl} mediate dark matter self-interactions which can significantly alter astrophysical signals compared to the standard cold dark matter paradigm. Self-interacting dark matter and these associated signals have been extensively studied (see Ref.~\cite{Tulin:2017ara} for a recent review article with extensive references therein). The self-interaction cross section as a function of velocity can be obtained by observations of systems at various scales, since the typical dark matter velocity in a galaxy cluster is much larger than in a dwarf galaxy~\cite{Feng:2009hw,Buckley:2009in,Loeb:2010gj,Kaplinghat:2015aga}.
The most stringent constraint arises from cluster scales where the cross section limits are $\sigma/m\sim\mathcal{O}(0.1)$ $\rm{ cm^{2}/g}$. On dwarf galaxy scales, the constraints relax to $\sigma/m\sim\mathcal{O}(1)$ $\rm{ cm^{2}/g}$.
Given a microphysical model, we can then compute this cross section, including non-perturbative Sommerfeld enhancement, and ensure that it satisfies the constraints at various scales, following the procedure in Ref.~\cite{Agrawal:2020lea,Parikh:2020ggm}. In our model, we are interested in mediators which are heavy relative to the dark matter, so the leading contribution to the scattering cross section is dominant and resumming the full Sommerfeld ladder does not cause significant deviations to the cross section. In particular, for our scenario, we have a symmetric population of dark matter, which in turn implies that both $\chi\chi\to\chi\chi$ and $\chi\chi^{\dagger}\to\chi\chi^{\dagger}$ are active. Both the free and Co-SIMP parameters contribute to these processes at tree- and loop-level. The scattering cross section limits then constrain only this combination of parameters, but not the Co-SIMP parameters themselves. In particular, we are free to choose a value of the Co-SIMP parameters which gives us a viable cosmology since the free parameters can be shifted accordingly to evade self-scattering constraints.

%%%%%%%%%%%%%%%%%%%%%%%%%%%%%%%%%%%%%%%%%%%%%%%%%%%%%%%%%%%%%
\subsection{Direct Detection Experiments}

A major testing ground for DM-SM interactions is the arena of direct detection (DD) experiments, where the program of large volume liquid noble gas detectors has achieved exquisite sensitivity for DM-nucleon interactions in the GeV--TeV range.  In recent years, the expansion of sensitivity towards sub-GeV dark matter candidates has become a key focus of DD experimental efforts, and rapid developments in low-threshold detector technologies have been made in response. The establishment of a set of benchmark models targeted by these future experiments has likewise become an area of intense study.  

From this perspective, the thermal relic Co-SIMP is an intriguing scenario to investigate. Not only are the expected scattering cross sections highly predictive as the interaction strength is set by the relic abundance, but the Co-SIMP interaction also leads to multiple types of signals that direct detection experiments may simultaneously search for.  Our model predicts not only both elastic and inelastic scattering signatures, but the additional possibility of elastic signatures from high-energy DM produced from local Co-SIMP interactions as well. We discuss each of these in turn in the following subsections.

\subsubsection{Elastic Scattering}
\begin{figure}
    \caption{
   These are the various two-loop elastic scattering diagrams for Co-SIMPs. While other diagram topologies exist at tree level and one loop, the leading contribution from the Co-SIMP parameters appears at two loops.
    }
    \resizebox{\columnwidth}{!}{
    \includegraphics[scale=0.45]{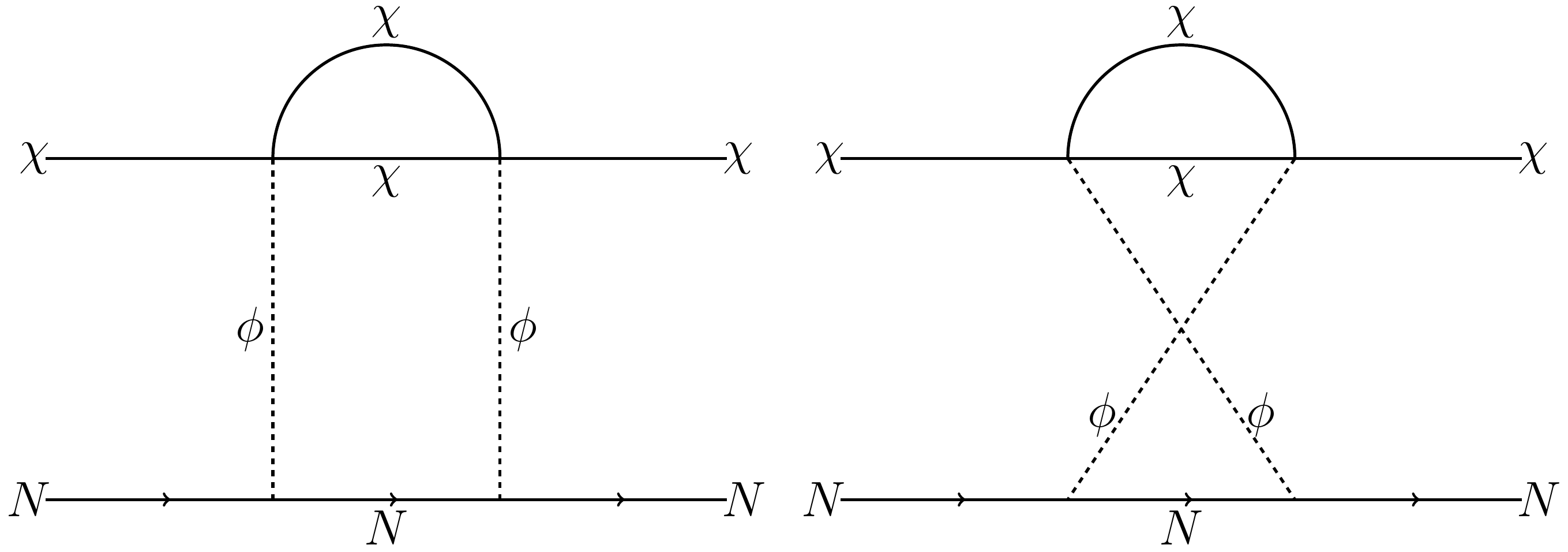}
       }
\label{fig:elastic_scattering_nucl}
\end{figure}

The elastic scattering process for our scenario is absent at tree-level and appears only at the two-loop level. In Fig.~\ref{fig:elastic_scattering_nucl}, we show the diagrams which lead to elastic nucleon-DM scattering for Co-SIMP dark matter. The diagram can be evaluated in two steps. First, we consider the $\chi$ loop, which renormalizes the $ \lambda_{\chi\phi}$ coupling of the operator $|\chi|^{2} \phi^2$, which we denoted $\lambda_{2}$ in Eq.~\ref{eqn:scalar_potential}. Given that it generates a coupling of the magnitude of the finite part of the loop integral, we find that $\lambda_{\chi \phi} \approx y_{\phi \chi}^2/(4 \pi^2) \equiv \alpha_D/\pi$. The dependence on the $\chi, \phi$ mass ratio is weak when $m_\phi > m_\chi$. 
Thus, the full loop induced spin-independent cross section can be calculated and is given by
\begin{align}
    \sigma_{\chi N} &  = \frac{ m_N^2 C_{\rm loop-q}^2 }{ 32 \pi \, \left( m_N  + m_\chi\right)^2},  \quad \text{  with } \\ \nonumber 
      C_{\rm loop-q} & = \sum_q \frac{\lambda_{\chi \phi} \,  y_{\phi q}^2 f_q^2 \left(  R \log
   \left(\frac{m_q^2}{m_\phi^2}\right)+ \left(1-2
   \frac{m_q^2}{m_\phi^2}\right) \log \left(\frac{m_\phi^2 (R+1)^2}{4
   m_q^2}\right)\right)}{8 \pi^2 \, m_q R},
\end{align}
with $R \equiv \sqrt{1 -4 m_q^2/m_\phi^2}$. In the limit of $m_\phi \rightarrow \infty$, the loop factor apporaches $C_{\rm loop} \approx (\alpha_D/\pi)  \sum_q y_{\phi q}^2 f_q^2/(4 \pi^2) \,m_q/m_\phi^2 $, as expected in the EFT limit. For our \textbf{Model I}, this coupling can be further simplified as 
\begin{align}
    \sum_q y_{\phi q}^2 f_q^2 m_q  \approx (y_{N}/16)^2 \left( f_u^2 m_u + f_d^2 m_d \right) \,.
\end{align} 
Given this process, a number of current and upcoming direct detection experiments have sensitivity to the nucleophilic Co-SIMP space. 

In Fig.~\ref{fig:nucleoparameterspace}, we summarize the limits and sensitivities from existing and planned experiments. There are three major classes of experiments. Large noble gas detectors, such as LZ~\cite{LZ:2022ufs} are insensitive to small recoil energy deposits for the sub-GeV masses. However, they can detect a fraction of DM particles that have been accelerated by cosmic rays~\cite{Cappiello:2018hsu,Bringmann:2018cvk}. Then, above ground experiments, which only have atmospheric overburden, test the largest values for the elastic cross section using solid-state detector technology~\cite{CRESST:2017ues}. Finally, a very promising novel technology based on superfluid helium will lead to detectors with a lowered energy threshold and efficiently probe large fractions of the nucleophilic Co-SIMP parameter space~\cite{Ruppin:2014bra,Schutz:2016tid,Hertel:2018aal}.  We also present this information in Fig.~\ref{fig:nucleosigmael}, displaying instead our predictions from a thermal relic Co-SIMP candidate in the space of elastic scattering cross sections, relative to the regions probed by these current and future experiments. Predictions such as these may serve as benchmarks on which to target searches for Co-SIMP dark matter in the future.

Finally, we stress that we consider here only the minimal interaction strength that is induced by the two-loop diagram, and is directly linked to the set of parameters that is determining the Co-SIMP relic abundance. 
Thus, this is the minimal, irreducible, elastic scattering cross section in this realization that is shown in the parameter space. Note, however, that larger cross sections are always possible and can be induced at tree-level by turning on free parameters in the Lagrangian. Thus, other exploratory searches, such as exoplanet heating~\cite{Leane:2020wob}, or searches based on DM surface abundance enhancement~\cite{Leane:2022hkk} are still very promising for testing the Co-SIMP scenario. A detailed analysis of scenarios that go beyond the minimal viable cross section values is left to future work.

\subsubsection{Monoenergetic Recoil from Inelastic Scattering}

A unique opportunity for experimental searches for Co-SIMPs is the possibility that the signature $\chi\chi N \to \chi^\dagger N$ interaction occurs within the detector itself. One of the big challenges of detecting sub-GeV DM particles is the fact that the energy deposit in the detector is correspondingly low and thus very low detection thresholds are needed. The Co-SIMP reaction, being strongly exothermic, does not require such low thresholds. The signature is a monoenergetic recoil at  $E_R \approx 1.5 m_\chi^2/m_{\rm SM}$ for the case of $m_\chi \ll m_{\rm SM}$.  The Co-SIMP reaction rate in a volume $V$ of SM particles is given by 
\begin{align}
    \Gamma_{3 \to 2} =  \int_V  \gamma_{\rm 3\to 2} d^3\vec{r}=  \int_V n_\chi^2 n_{\rm SM} \langle \sigma v^2\rangle d^3\vec{r} \,.
    \label{eq:32rate}
\end{align}

The benchmark interaction density for a thermal relic Co-SIMP is then predicted to scale as 
\begin{align}
    \gamma_{3 \to 2} \simeq \frac{0.1}{ \text{m}^3\,  \text{day}} \left(\frac{n_{\rm SM}}{N_A \text{ cm}^{-3}}\right) \left( \frac{\rho_\chi}{0.4 \; \text{GeV}/\text{cm}^3} \right)^2  \left(\frac{0.1 \, \text{MeV}}{ m_\chi}\right)^5 \,,
\end{align}
where $N_A$ is Avogadro's number, and the cross section has been chosen to satisfy $\Omega_\chi h^2 =0.12$ for $m_\chi = m_\phi$.  While the event rate compared to that of elastic scattering is suppressed by another power of the halo DM abundance, the cross section governing this process is present at tree level and much larger. However, while the energy deposit for this process evades the suppression of low halo velocity, it is often suppressed by the mass ratio of the DM particle to the much heavier SM nucleus. Thus, much of our parameter space is expected to be below threshold for a Xenon-type experiment.  On the other hand, more futuristic technologies such as superfluid Helium may be very effective at probing our parameter space due to the lighter nuclei mass and significantly lowered detection thresholds. It is particularly compelling then, that such a future detector should expect to see both elastic and inelastic scattering signals from our model.

\subsubsection{Boosted Co-SIMPs and Accumulation in the Earth}

A third type of possible signature, the synthesis of the two types of interactions described above, is found in the elastic recoil of a boosted DM particle from the Co-SIMP interaction.  That is,  if a Co-SIMP reaction takes place in the material surrounding the detector or within the Earth, the relativistic reaction product may be  detected via an elastic collision with the SM particles in the detector.

The expected flux of boosted Co-SIMP particles on the Earth's surface is given by $\Phi_{\oplus} = \Gamma^{\oplus}_{3\to 2}/A_{\oplus}$, where $\Gamma^\oplus_{3\to 2}$ is given by Eq.~\ref{eq:32rate} and integrated over the volume of the Earth. Naturally, this rate is sensitively dependent on the density of DM within the Earth available to source this interaction. It has been shown that in addition to the halo DM component, a captured DM population can be present in the planet if elastic DM-SM interactions are present~\cite{Leane:2020wob,Gould:1988ym,Gould:1988eq,Gould:1989hm}.  In particular, recent results~\cite{Leane:2022hkk, Acevedo:2023} show that the minimal DM mass that is retained in a celestial body can be substantially lower than previously expected, and even MeV-scale DM can accumulate inside the Earth. 

Assuming that the Earth can be approximated as a homogeneous body, we can simplify our results and find
\begin{align}
    \Gamma_{\rm 3\to 2} \approx  \frac{N_\chi^2  \langle \sigma v^2\rangle n_{\rm SM} }{V_{\rm eff}} \quad \text{   where    } \quad V_{\rm eff} = \frac{\left( \int_V n_\chi r^2 dr \right)^2 }{\int_V n_\chi^2  r^2 dr }  \,,
\end{align}
and $N_\chi$ is the total number of Co-SIMP particles in the planet. The spatial distribution of the DM depends on the interaction regime. For sufficiently large scattering cross sections, local thermal equilibrium (LTE) is achieved, leading to a profile strongly affected by diffusion~\cite{Gould:1988ym,Leane:2022hkk}. We however assume that our elastic interactions do not induce the LTE regime and instead use the isothermal distribution, which can be simply derived from the hydrostatic equilibrium condition. The normalized distribution reads 
\begin{align}
    n_\chi(r) = n_0 \,  e^{- \frac{m_\chi}{T_\chi} \phi(r)}\, ,
\end{align}
where $\phi(r)$ is the gravitational potential inside the object, $T_\chi$ is the DM temperature, which is approximately 90\% of the core temperature of the object~\cite{Spergel:1984re}, and $n_0$ is a normalization factor set by the condition $ \int_V n_\chi dV  = N_\chi$. 

This number of dark particles $N_\chi$ is in turn set by the gravitational capture of ambient DM due to DM-SM scattering. We denote the capture rate $\Gamma_{\rm cap}$. It can be parametrized by the DM flux through the planet times a capture efficiency $f_{\rm cap}$, which has been derived in Ref.~\cite{Leane:2022hkk}, including the reflection correction for light DM. Thus, the capture rate is $\Gamma_{\rm cap} = f_{\rm cap} \phi_{\chi} \pi R_{\oplus}^2$, where we have neglected the effect of gravitational focusing, and the effect of the motion with respect to the DM halo, as they are subdominant for the Earth. The DM flux is given by $\phi_{\chi} =  v_{\chi} \rho_{\chi}/m_{\chi}$, with $\rho_{\chi}$ being the DM halo density. The accumulated DM abundance is then given by $N_\chi = \text{min}\{ N_{\chi}^{\rm eq} , \, \tau_{\oplus} \Gamma_{\rm cap} \}$. Here the equilibrium value $N_\chi^{\rm eq}$ is derived from the condition $\Gamma_{\rm 3\to 2} = \Gamma_{\rm cap}$ and is given by
\begin{align}
    N_\chi^{\rm eq} = \left(  \frac{ \Gamma_{\rm cap} V_{\rm eff}}{n_{\rm SM} \langle  \sigma v^2 \rangle} \right)^{1/2} \,. 
\end{align}

Under the assumption that Earth has entered the depletion-capture equilibrium, the flux of outgoing accelerated Co-SIMPs can be written as 
\begin{align}
   \Phi_{\oplus} = \frac{\Gamma^{\oplus}_{3 \to 2}}{A_{\oplus}} = \frac{f_{\rm cap} \phi_{\chi} \pi R_{\oplus}^2}{A_{\oplus}} = \frac{f_{\rm cap}  \, \rho_\chi^{\rm halo} v_{\chi}}{4 m_\chi}\,. 
\end{align}
We assume that a detector with mass $M_T$ and a target material with atomic mass $m_A$ has an energy threshold that allows sensitivity to recoil events at the accelerated Co-SIMP momenta
\begin{align}
    \frac{p_{\rm acc}}{m_{A}} = \frac{M_R \sqrt{ 3 \left(M_R + 2 \right)\left(3 M_R + 2 \right) } }{4 M_R + 2} \approx \sqrt{3} \left( M_R - M_R^2 \right) \quad \text{   with    } \quad M_R = \frac{m_\chi}{m_{A}}  \,.
\end{align}
Given the induced Co-SIMP-nucleus cross section $\sigma_{\chi A}$, coherently enhanced relative to the DM-nucleon scattering $\sigma_{\chi N}$, the event rate in the detector is 
\begin{align}
    \Gamma_{{\rm boosted} \; \chi A} = 1.35 \, \frac{f_{\rm cap}}{4} \left(\frac{ v_\chi}{240 \, \text{km}/\text{s}}\right)  \left(\frac{ \rho^{\rm halo}_\chi \, \text{cm}^3 }{ m\chi} \right)  \left(\frac{ \sigma_{\chi A} }{ \text{pb}}\right)
     %\left(\frac{ M_T \, \text{GeV} }{ m_A \, \text{ton} }\right) 
     \left(\frac{ M_T }{ \text{ton} }\right) 
     \left(\frac{ \text{GeV} }{ m_A }\right) 
     \frac{1}{\text{s}} \,.
\end{align}
It is additionally intriguing to note that the dependence on the Co-SIMP interaction cross section cancels from this prediction, encapsulated in the assumption that the DM accumulation is set by the equillibrium condition. In Sec.~\ref{sec:nucelo_param_space}, we explore the sensitivity of current and planned experiments to this signature. The fact that the Co-SIMP framework predicts three different signals in direct detection experiments -- namely the standard halo low-threshold elastic scattering, the accelerated Co-SIMP high-threshold elastic scattering, and the monoenergetic recoil from direct Co-SIMP conversion process -- makes this scenario particularly salient to study in the arena of DD experiments. The predictions of these three signals, correlated in strength, articulate in total a smoking-gun signature for Co-SIMP dark matter.

\subsection{Other Terrestrial Searches}
In addition to searches at dedicated dark matter experiments, other particle physics experiments may also potentially be critical for discovering or ruling out our model. Indeed, the introduction of a light scalar mediator $\phi$ that couples to baryons in this UV completion opens up several avenues of possible detection. In the following, we present a number of experimental opportunities and discuss their short-term optimization potential.

%%%%%%%%%%%%%%%%%%%%%%%%%%%%%%%%%%%%%%%%%%%%%%%%%%%%%%%%%%%%%
\subsubsection{Meson decay}

\begin{figure}
    \caption{These are the diagrams which generate loop induced meson decay. The left diagram generates $B\to K\phi$ decays and the right diagram generates $K \to \pi\phi$ decays. Since these amplitudes are proportional to the mass of the quark running through the loop, the dominant contribution arises from top-$W$ loops.
    }
    \resizebox{\columnwidth}{!}{
    \includegraphics[scale=0.45]{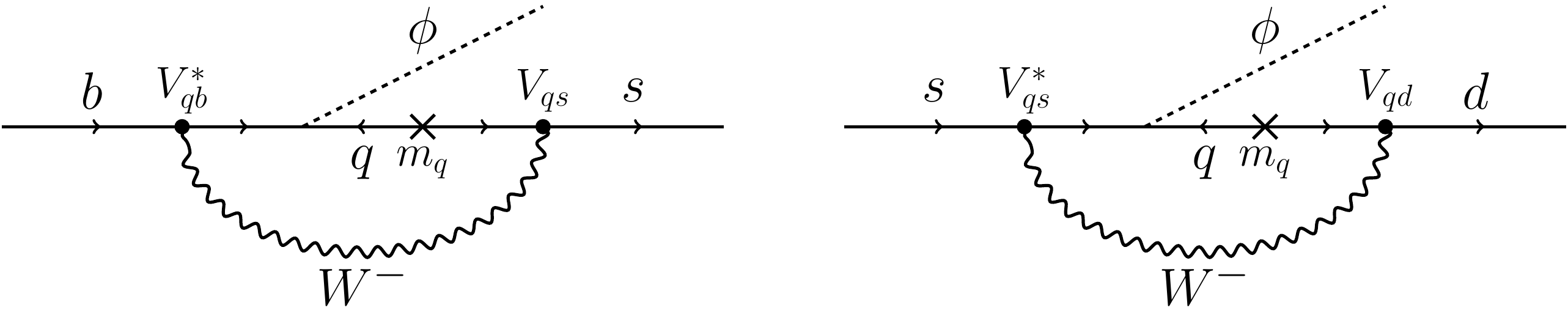}
       }
\label{fig:meson_decay}
\end{figure}

Light scalar particles with nucleophilic interactions could facilitate rare meson decays, such as $B \rightarrow K \phi$ and $K \rightarrow \pi \phi$, as shown in Fig.~\ref{fig:meson_decay}. Those processes are dominantly driven by top-$W$ loop contributions in the SM, and thus the branching ratios into $\phi$s are set by the top-$\phi$ interaction. The contributions of first-generation quarks to these meson widths are severely suppressed by their small Yukawa couplings. 

As discussed in Ref.~\cite{Knapen:2017xzo}, the most stringent constraints for scalars lighter than the pion come from $K \rightarrow \pi \phi $ decays, which leads to a $K \rightarrow \pi$ and missing energy signature in all of the relevant Co-SIMP parameter space. The experimental observation of the branching ratio $\text{BR}(K \rightarrow \pi \bar{\nu} \nu ) = 1.73^{+1.15}_{-1.05} \times 10^{-10}$~\cite{E949:2008btt} thus limits the top quark coupling to $y_t <  7.9 \times  10^{-5}$. However, contribution from the lighter quarks to the decay rate scales as their mass times respective coupling constants squared, so bounds on light quark couplings are dramatically weaker.  If all quark couplings saturate the experimental meson decay bound, the effective coupling to nucleons is constrained to be at most $y_N^{\rm max} = \sum_q f_q y_q^{\rm max} \approx \mathcal{O}(10)$, which is greatly disfavored by direct detection constraints.

%However, as we discusses above the top quark coupling is not required to be large for Co-SIMP nucleon interactions to be effective.

%    \js{top quark coupling has to be avoided. Potentially coule to lightest gen. only, or even just u-quark...}

%%%%%%%%%%%%%%%%%%%%%%%%%%%%%%%%%%%%%%%%%%%%%%%%%%%%%%%%%%%%%
\subsubsection{Beam Dump Experiments}
% intro
Experiments located at beam dump facilities have a strong discovery potential for light dark sectors that involve particles with long lifetimes. We investigate their reach towards direct tests of the nucleophilic Co-SIMP interaction. As an example, we consider a recent search performed at Fermilab, where an 8 GeV proton beam was directed at the beam dump, and the MiniBooNE experiment was the detector shielded by $\sim 500$ meter of earth material~\cite{MiniBooNE:2017nqe, MiniBooNEDM:2018cxm}.  
\begin{figure}
    \caption{
   Diagrams of the two processes which can produce a signature that can be explored in beam dump experiments. Left panel: production process via an off-shell mediator. Right panel: inelastic process, that would lead to ionization signatures in the detector. This process can only proceed as long as $\chi$ particles have sufficient energy. 
    }
    \resizebox{0.8\columnwidth}{!}{
    \includegraphics[scale=0.45]{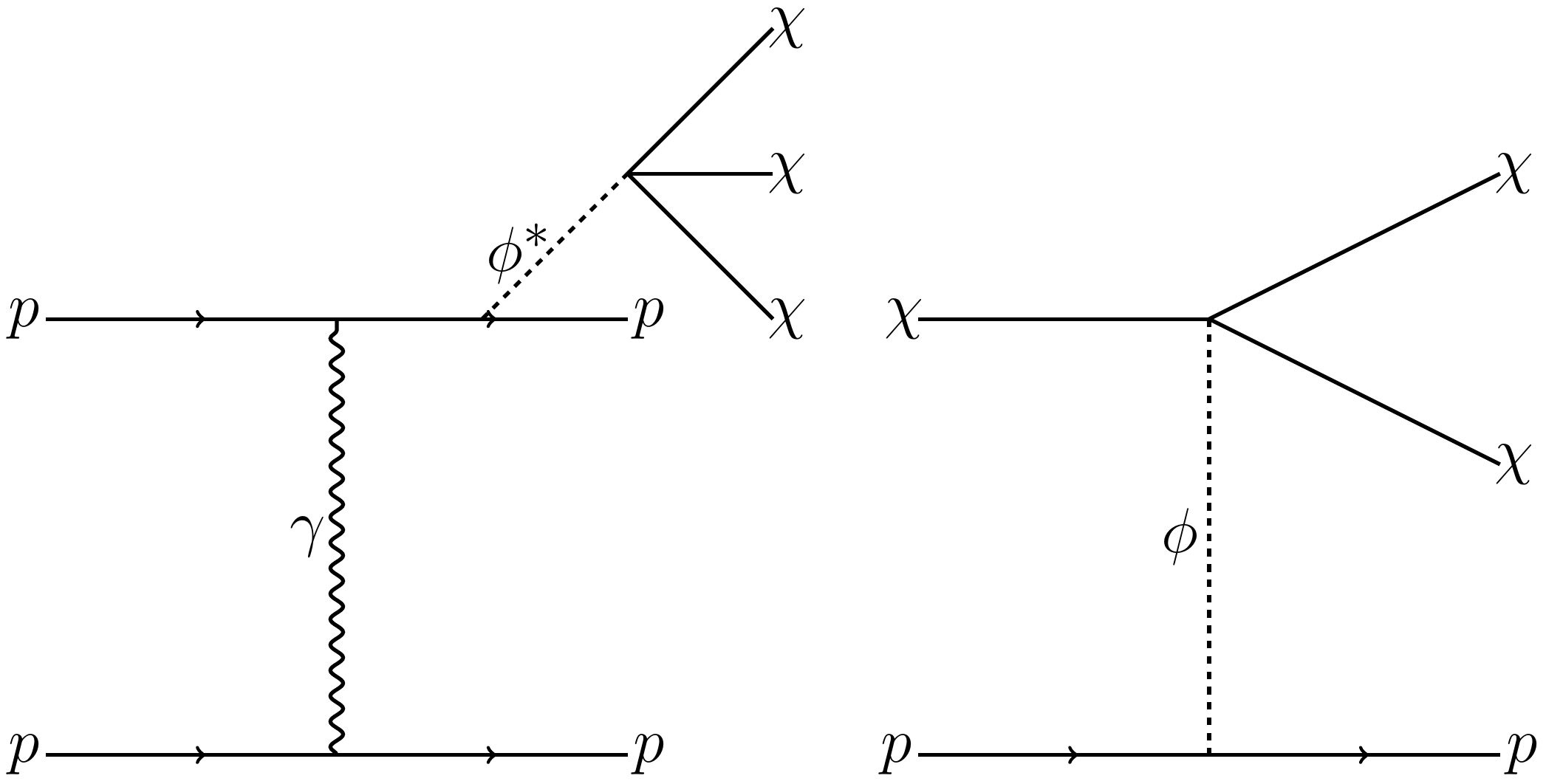}
       }
\label{fig:nucleo_beam_dump_diagrams} 
\end{figure}

% production and detection
In the left panel of Fig.~\ref{fig:nucleo_beam_dump_diagrams}, we show the production process for Co-SIMP DM, which is based on radiating a $\phi$ mediator during proton scattering $p + p \rightarrow p + p + \phi^*$. Since this is an off-shell process, with excess energy, the mediator produces relativistic Co-SIMPs via $\phi^* \rightarrow 3 \chi$. The cross section for this process scales as $\sigma_{\rm prod} \sim \alpha_{\phi N} \alpha_{D} \sigma_{pp \,  \rm elastic}$. As the Co-SIMPs are produced with an energy excess, they can undergo inelastic scattering with SM nucleons via $\chi + p \rightarrow \chi + \chi + p$, where a t-channel $\phi$ is exchanged. This process, shown in the right panel of Fig.~\ref{fig:nucleo_beam_dump_diagrams}, has a cross section which scales as $\sigma_{\rm det} \sim \alpha_{\phi N} \alpha_{D} m_{\phi}^{-2}$.

In the parameter space of interest, the mean free path of Co-SIMPs that are energetic enough to inelastically undergo the above cascade reaction varies on the scale $\lambda_{\rm mfp} \sim 10^{-2} - 50 \, \text{meters}$, with the larger mean free path for $\phi$ masses above the GeV scale. Thus, a few Co-SIMPs could react in the MiniBooNE detector volume in only a fraction of the parameter space.

In Fig.~\ref{fig:nucleoparameterspace}, we show the region in the parameter space where the results of Refs.~\cite{MiniBooNE:2017nqe,MiniBooNEDM:2018cxm} have potential sensitivity to the Co-SIMPs. We see that mass and coupling values predicted by the thermal production can be probed by a dedicated search for the proposed signature. As the mean free path of the events deviates from the one expected for SM interactions, the events would not look like an SM particle depositing energy in the detector. Thus, a dedicated analysis investigating such exotic ionization signatures is highly motivated. 

A possible way to expand the beam dump experimental sensitivity in the future is to reduce the shielding distance between the production and detection region. Furthermore, dedicated collider searches for dark shower events, with a variable step size, related to the $\lambda_{\rm mfp}$ of Co-SIMPs would be required to move deeper into the parameter space with lighter mediators and shorter scattering interaction lengths.

%
%\js{ $N + N \rightarrow \phi \rightarrow \gamma \gamma$}
%\js{proton on target beams: MiniBoon:  https://arxiv.org/pdf/1702.02688.pdf}
%\js{$\phi \rightarrow \gamma \gamma $ suppressed as $m_eff$ in loop small}

%\ap{Beam dump produces a phi with energy sqrt{Ebeam*mnucleon} which is generically larger than mphi. If mphi > 3mchi then you immediately decay into chis and then you just have to worry about the chi propagating to the detector and scattering via chi + nucleon -> chi + chi + nucleon. The DD process will always be subdominant since the chi will always be super relativistic and have enough energy to create a second chi.
%If mphi < 3mchi, then you can propagate to detector as a phi and then either produce two photons or scatter phi with nucleon leaving a recoil signature.}

%\ap{add a few sentences here about how these bounds show up in Fig. 5}

\subsection{Parameter Space}
\label{sec:nucelo_param_space}

\begin{figure}
\caption{ Parameter space of our nucleophilic Co-SIMP model with a scalar mediator. \textit{Shaded regions} show current constraints~\cite{Grifols:1986fc,1989MPLA....4..311G,CRESST:2017ues,Cappiello:2018hsu} and the \textit{dashed lines} shows projected sensitivities. Specifically, the \textit{dark blue line} represents the superfluid He sensitivity~\cite{Ruppin:2014bra, Schutz:2016tid, Hertel:2018aal} to the elastic scattering process and the \textit{green line} represents the superfluid He sensitivity to the $3 \rightarrow 2$ Co-SIMP process. The \textit{orange line} shows the sensitivity of a dedicated search with the MiniBooNE beam dump experiment~\cite{MiniBooNE:2017nqe,MiniBooNEDM:2018cxm}, and the \textit{magenta line} shows a search for boosted Co-SIMPs from inside the Earth with a liquid noble gas detector~\cite{Aalbers:2022dzr}. The \textit{dark red lines} corresponds to the parameter space that gives the correct present-day relic abundance, with various choices of $\mu_R\equiv m_\phi/m_\chi$. In this parameter space, fifth force experiments~\cite{Murata:2014nra} and low energy neutron scattering searches~\cite{Leeb:1992qf,Pokotilovski:2006up,Nesvizhevsky:2007by,Kamiya:2015eva} are not constraining as they are only competitive at mediator masses lighter than those we consider here.}
    \resizebox{0.6\columnwidth}{!}{
    \includegraphics[scale=0.45]{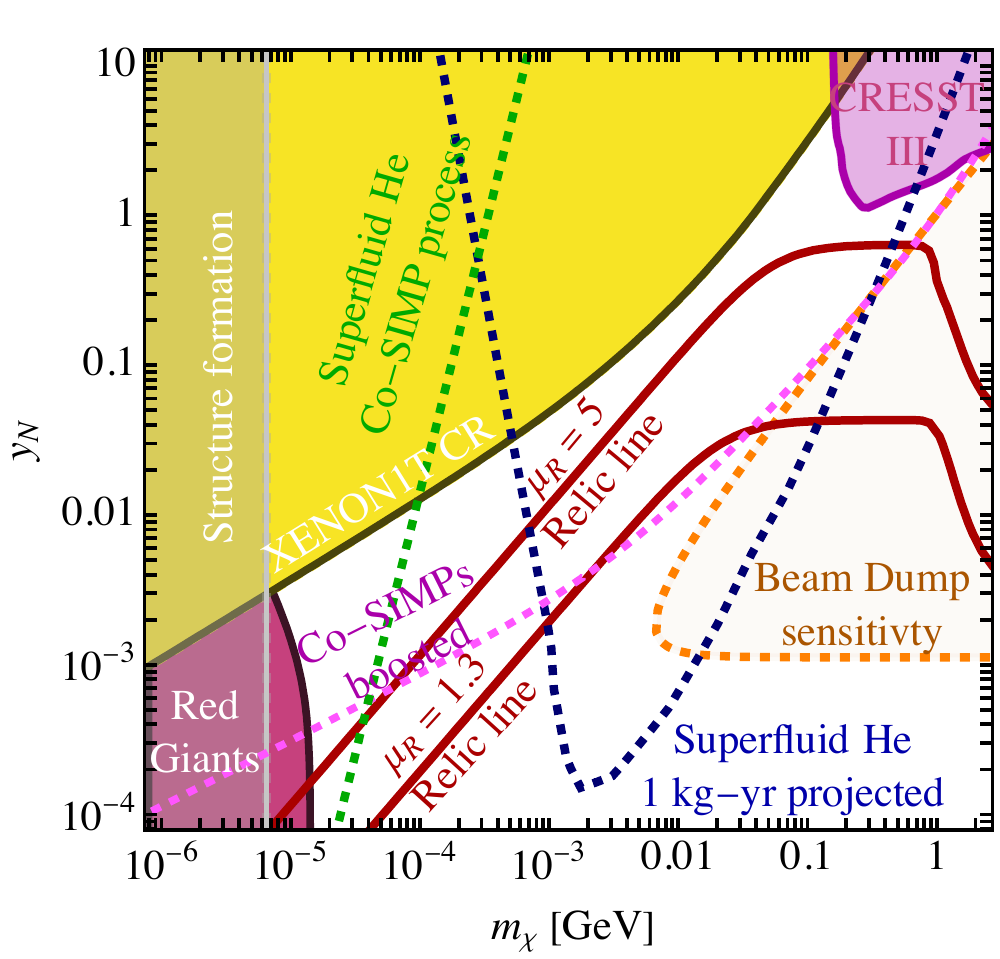}
       }
     \label{fig:nucleoparameterspace}
\end{figure}

In this section, we discuss the current bounds on the Co-SIMP parameter space and the sensitivities of upcoming experiments. We also show the potential of dedicated searches, using existing data, which could be reinterpreted under the Co-SIMP model hypothesis. Filled solid contours are chosen to represent existing bounds while dashed contours indicate the expected reach of new searches. 

Fig.~\ref{fig:nucleoparameterspace} shows the full parameter space of the scalar-mediated nucleophilic Co-SIMP model, including the curves producing the correct DM relic abundance for two choices of the mass ratio $\mu_R \equiv m_\phi/m_\chi = 1.3$ and $\mu_R = 5$.  The lines and contours shown in the DM mass-Yukawa coupling ($m_{\chi}$-$y_N$) plane assume that the DM-mediator coupling $y_{\phi \chi}=4\pi$ is maximal. The maximal value is defined by requiring unitarity on the cross section of the pure dark sector process $4 \chi \rightarrow 2 \chi$. In practice, we do this by requiring that the cross section in Eq.~\ref{eq:crosssectionsNucleo} satisfy the constraints derived in Ref.~\cite{PhysRevLett.103.153201,Asadi:2021bxp}. Thus, the values of the Yukawa coupling to the SM are minimal and we can demonstrate the full current and expected sensitivity reach to the Co-SIMP parameter space.

The filled contours show excluded regions in the parameter space. 
The direct detection limits of current low threshold experiments such as CRESST probe only the high mass end of our considered space and have not reached the thermal relic predictions yet~\cite{CRESST:2017ues}.
A significant fraction is covered by the XENON1T analysis, where a fraction of dark matter has been boosted above the XENON1T detection threshold by collisions with cosmic rays~\cite{XENON:2017vdw,XENON:2018voc,Cappiello:2018hsu}. The low mass end is covered by structure formation bounds~\cite{Chatterjee:2019jts} and bounds from the anomalous cooling of red giants facilitated by mediator emission~\cite{Grifols:1986fc,1989MPLA....4..311G}. BBN can test Co-SIMP masses below $10$ MeV, but the sensitivity is severely model dependent as we discuss in the next section, so we omit these constraints from our discussion here; note here simply that light Co-SIMP realizations would require additional model-building. 

The dashed contours show the expected future sensitivity of dedicated Co-SIMP searches. An analysis of MiniBooNE~\cite{MiniBooNE:2017nqe,MiniBooNEDM:2018cxm} data, as discussed in the beam dump section above, could test the heavier Co-SIMP mass range at small coupling values. The expected sensitivity is shown as the dashed orange contour. The dashed magenta contour indicates the sensitivity of a new analysis of 1 ton-years of XENON1T data, or data of a ton-year exposure of a similar experiment~\cite{Aalbers:2022dzr} searching for elastic signals for $\chi$ particles that have been boosted by the Co-SIMP interaction inside our planet. We expect that an event rate which can lead to up to 100 events per year in the detector volume is detectable. Overall, the search for a flux of boosted Co-SIMPs from inside our planet could greatly benefit from a large detection volume, as in the KamLAND-Zen~\cite{KamLAND-Zen:2014hhc} or JUNO~\cite{JUNO:2015zny} experiments.

Finally, a one kg-year search using a new low threshold detector technique, based on superfluid Helium~\cite{Ruppin:2014bra, Schutz:2016tid, Hertel:2018aal}, will cover a significant region of the parameter space. 
The sensitivity to the elastic scattering process is shown by the dashed blue contour, while the green contour shows sensitivity to the direct $3 \rightarrow 2$ Co-SIMP interaction. It dominates the sensitivity at low Co-SIMP masses since the event rate scales quadratically with DM density. For the superfluid helium sensitivity, we have assumed only the Co-SIMP halo number density, which is a conservative choice. As we discussed in a previous section, in scenarios that allow Co-SIMP accumulation in the Earth, those sensitivities will be greatly enhanced. 
\begin{figure}[h!]
    \caption{Benchmark cross section values for elastic scattering in the nucleophilic Co-SIMP scenario, predicted by the DM production process. The \textit{filled solid contours} show the sensitivity of current direct detection searches to elastic scattering induced by the Co-SIMP flux from the DM halo. Current low threshold searches~\cite{CRESST:2017ues}, shown in the \textit{magenta region}, do not reach the benchmark cross section values shown by the \textit{dark red lines}. The subdominant flux of Co-SIMPs boosted by cosmic rays provides a limit at the largest elastic cross sections~\cite{Cappiello:2018hsu}, and are shown as the \textit{yellow filled region}. Finally, the sensitivity of a new detector based on superfluid helium~\cite{Ruppin:2014bra, Schutz:2016tid, Hertel:2018aal} is shown as the \textit{blue dashed contour}.  
    }
    \resizebox{0.6\columnwidth}{!}{
    \includegraphics[scale=0.45]{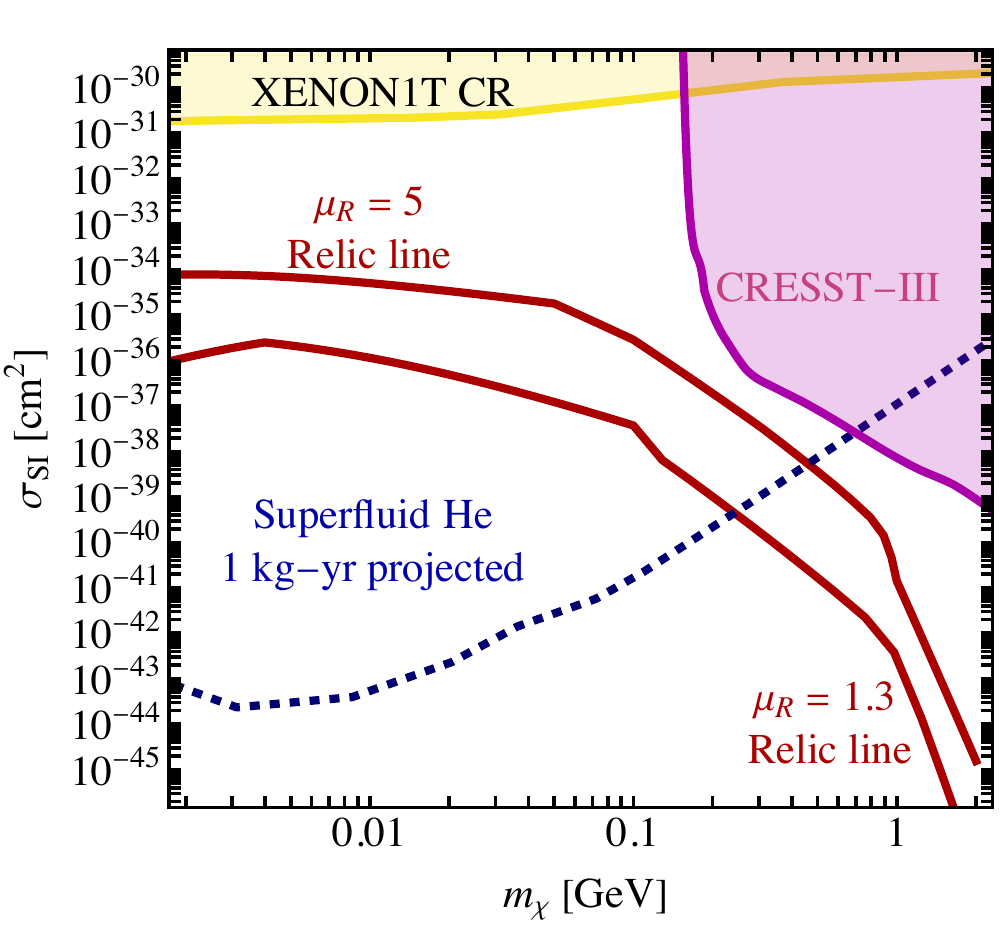}
       }
     \label{fig:nucleosigmael}
\end{figure}

Fig.~\ref{fig:nucleosigmael} shows the range of induced elastic cross sections given the Co-SIMP parameters predicted by the thermal freeze-out. The current generation of low threshold experiments~\cite{CRESST:2017ues} have not reached these cross section values yet, but future low threshold technologies are expected to largely cover this parameter space~\cite{Ruppin:2014bra, Schutz:2016tid, Hertel:2018aal}. In Fig.~\ref{fig:nucleosigmael} we only show sensitivities of experiments that search for the elastic direct Co-SIMP signal from the Co-SIMP halo flux, clearly defining this new experimental benchmark scenario and the predictions of a thermal cosmology within it.

Overall, we emphasise that in addition to the induced elastic Co-SIMP signatures from the halo flux at recoil energies of $E_R \sim 10^{-6} \, m_\chi$, there could be two more direct detection signals. One is the Co-SIMP reaction that leads to monoenergetic nuclear recoils at $E_R \sim 1.5 \, m_\chi^2/m_N$. The other is an elastic scattering signal from Co-SIMPs that have been boosted by the Co-SIMP number changing reaction to kinetic energies of the order of $m_\chi$ inside our planet. Thus, the Co-SIMP scenario has a unique property in the sense that it leads to three different direct detection signatures that are linked to the $3 \rightarrow 2$ reaction process which sets the relic abundance in the early universe, leading to greater testability of this production process in direct laboratory experiments.

\section{Leptophilic Co-SIMPs}
\label{sec:scalar_lepto}

Another simple scenario we can consider is one where the mediator couples the dark matter to electrons. Electrophilic Co-SIMP dark matter generically requires lower masses in order to achieve efficient enough scattering to freeze-out to the correct relic abundance. These lighter candidates are potentially very interesting for near future direct detection experiments where the technology to resolve low-threshold events is rapidly maturing. At the same time, DM in these mass ranges are subject to much stronger cosmological and astrophysical constraints. In this section, we examine the scenario of light leptophilic Co-SIMPs, delineate current constraints, and discuss the near-future detection possibilities. 

\subsection{Model setup}

As in the previous section, we assume that both the DM and mediator are new scalars. The Lagrangian is given by
\begin{align}
\mathcal{L} &\supset  \partial_\mu \chi^\dagger \partial^\mu \chi +  \frac{1}{2} \left(\partial_\mu \phi\right)^2  -   m_\chi^2 |\chi|^2  -  \frac{1}{2} m_\phi^2 \phi^2 -  y_{\phi e} \phi \bar e e - \frac{y_{\phi\chi}}{3!} \chi^3\phi  - V(\chi,\phi)  
\label{eqn:scalar_lagrangian_lepto}
\end{align}
where the general potential $V(\chi, \phi)$ is given by Eq.~\ref{eqn:scalar_potential} and we make the same notational distinction between ``free parameters" and ``Co-SIMP parameters". Mediators that couple universally to all lepton generations might just as easily be considered, but in practice all cosmological processes and experimental constraints are dominated by interactions with electrons. 

As before, $\phi$ is neutral and $\chi$ is charged under a   $\mathbb{Z}_{3}$ symmetry, and the  Co-SIMP interaction is given by the middle diagram of Fig.~\ref{fig:dm_depletion_nucl} with the nucleon line replaced by an electron line. As before, this process, as well as the corresponding process at 1-loop, naturally contribute to the depletion of dark matter abundance.
% UV coupling
%Note that the Co-SIMPs couple to charged leptons, but not to neutrinos. This is required in order to kinetically close channels with only two $\chi$ particles in the initial state which would otherwise lead to a WIMP-like freezeout.
Note that in these scenarios, Co-SIMPs may couple to charged leptons, but not to neutrinos. This is required in order to prevent the abundance being set instead by WIMP-like freeze-out via $\chi\chi\to\nu\nu$. 

In order to preserve the $SU(2)_L$ gauge symmetry in this model, additional ingredients are needed in the UV. We introduce a new vector-like fermion $\tilde{E}$, which allows an explicit mass term $m_{\tilde E} \bar{\tilde{E}} \tilde{E}$. For this vector-like fermion, $\tilde{E}_L$ and $\tilde{E}_R$  have the quantum numbers of the SM right-handed electron. This allows for an operator $ y_E  H \bar{L} \tilde{E}_{R}$, where $L$ is the SM lepton doublet and $H$ is the SM Higgs field, as well as the operator $y_{E_R} \phi \bar{\tilde{E}}_{L} e_R$,  and $\tilde{M} \bar{\tilde{E}}_{L} e_R$ where $e_R$ is the right-handed electron field. After we integrate out $\tilde{E}$ and break electroweak symmetry, the effective operator $y_E y_{E_R} \, (v_{H}/m_{\tilde{E}}) \, \phi \, \bar{e}_L e_R$ is induced. 

The coupling of the scalar mediator to SM charged leptons in Eq.~\ref{eqn:scalar_lagrangian_lepto} is given by $y_{\phi e} = y_E y_{E_R} \, (v_{H}/m_{\tilde{E}})$. Given the required model parameters, the mass of the $\tilde{E}$ field can be in the multi-TeV range, and thus not directly relevant to the Co-SIMP phenomenology we are investigating in this work and out of reach of collider searches for new charged states. Note that this construction also introduces a contribution to the electron mass, which scales as $\Delta m_e \sim \tilde{M} v_H y_{E}/M_{\tilde{E}}$, which given the parameter space of our model requires $\tilde{M} < 100 \, \text{GeV}$.

%\js{Discuss late thermalization

%\js{Discuss scalar couplings and issues of $Z_2$ realization and naturalness (small parameters that are needed, and how small they are).}

%\js{Show how coupling to electrons is possible without violating $SU(2)$. We had the UV-complete construction in the notes.}

%%%%%%%%%%%%%%%%%%%%%%%%%%%%%%%%%%%%%%%%%%%%%%%%%%%%%%%%%%%%%
%%%%%%%%%%%%%%%%%%%%%%%%%%%%%%%%%%%%%%%%%%%%%%%%%%%%%%%%%%%%%
\subsection{Early Universe Cosmology}

In the leptophilic scenario, the kinematic requirement that $m_\chi < 2\, m_e \sim \, \text{MeV}$ is compatible with BBN observations in the case of late thermalization. Thus, we assume that the same late thermalization mechanism, as detailed for the nucleophilic scenario, brings the dark sector and SM in equilibrium after the end of BBN, and the Co-SIMP abundance is populated afterwards. The late time relic abundance in this case is also set by the freeze-out of the Co-SIMP interactions, as detailed in Eq.~\ref{eq:boltzmann} with the nucleons replaced by electrons.

\subsection{Astrophysical Reach}
Theories with new, light, weakly-coupled particles in the dark sector that interact with the Standard Model and with each other have to contend with two major astrophysical constraints: dark matter self-scattering and stellar cooling bounds~\cite{Hardy:2016kme}. Since at leading order dark matter self-scattering processes do not involve any SM species, the leptophilic constraints are identical to those in the nucleophilic case and we refer the readers to the discussion in Sec.~\ref{sec:scalar_leptophilic_astrophysical_constraints}.
Rather, we focus the following discussion on stellar cooling bounds.

%%%%%%%%%%%%%%%%%%%%%%%%%%%%%%%%%%%%%%%%%%%%%%%%%%%%%%%%%%%%%
%\subsubsection{Stellar cooling bounds}

%An almost universal constraint that theories with new, light, weakly-coupled particles need to contend with is the production and escape of these particles in astrophysical environments.
%Stellar cooling constraints deal with the production and escape of new particles.
If the stellar core produces a hot enough thermal environment to support a population of light dark particles -- in our case the light mediator $\phi$ -- they could see the star as effectively transparent since they are weakly-coupled and escape. This carries away energy from the stellar interior, causing an anomalous cooling of the star and deviations from well-measured rates of stellar evolution.

In broad strokes, these constraints may be circumvented by making the particles too heavy for thermal production in the core (typically of temperatures of $\mathcal{O}(10 - 100 
\text{ keV})$ for red giants), or too short-lived to escape the radius of the star before decaying into (trapped) SM particles. The SM decay channel $\phi \to \gamma \gamma$ is computed in Eq.~\ref{eqn:phidecay} (applicable here with the substitution $e$ for $N$), but we additionally need to check here that the branching ratio is not in fact dominated by decays into the dark sector -- in that case, even decays that occur within the star do not prevent energy from escaping.  And indeed, the decay width of $\phi$ into dark matter, $\phi \rightarrow \chi\chi\chi$, is induced by a single vertex process and expected to dominate since the dark sector internal couplings are generically much stronger than those with the SM.  Therefore, it is useful to kinematically forbid this channel by imposing $m_\phi < 3 m_\chi$; in contrast, the SM decay to $2\gamma$ is always kinematically available, though loop-suppressed. The resultant stellar cooling bound, which constrains $\phi$'s that are too light or too long-lived, is shown in Fig.~\ref{fig_PS_scalarLep} in conjunction with additional constraints discussed in following subsections.

\subsection{Direct Detection Experiments}

In recent years, the sensitivities of direct detection experiments have dramatically increased in the realm of sub-GeV dark matter with electron interactions becoming the focus in this lighter regime. Likewise, this substantial step forward in experimental technology has greatly increased the appetite for the study of leptophilic  dark matter models that lie in this mass range. At this junction, the light leptophilic Co-SIMPs represent an interesting target for these future low-mass experiments to search for. 

The detection prospects are especially intriguing since the Co-SIMPs interact with the SM in both elastic and inelastic ways which creates a number of different signals at direct detection experiments.  The traditional elastic scattering with electrons is present, though in our model this process is mediated at 2 loops.  In addition, the tree level $3\to 2$ process can occur within the detector, which leads to a monoenergetic electron recoil signature.  Finally, the same process can also occur in the Earth, which can accelerate the DM enough for it to surpass the energy thresholds of the detector and induce energetic elastic scattering signatures.  We discuss these in more detail and estimate their rates in the subsequent discussion. The correlated predictions of multiple signals potentially in the same direct detection experiment also constitute a smoking-gun signal, allowing us to identify any future detections as definitively Co-SIMP or not. 

\subsubsection{Elastic Scattering}

First, one must of course consider the potential elastic direct detection signatures.  %Indeed the XENON1T collaboration has recently announced the discovery of a moderate excess in electron recoil data, which could be construed as evidence for DM-electron scattering at mass ranges similar to those considered here. 
The elastic scattering $\chi e \to \chi e$ in this Co-SIMP realization occurs at 2-loops, with the leading order diagrams given by replacing the nucleon line with an electron line in Fig.~\ref{fig:elastic_scattering_nucl}. The cross section is

\begin{align}
    \sigma_{\chi e} &  = \frac{ m_e^2 C_{\rm loop-e}^2 }{ 32 \pi \, \left( m_e  + m_\chi\right)^2},   \quad \text{  with } \\ \nonumber 
   % C_{\rm loop}  &  \approx \frac{\alpha_{\chi \phi} \alpha_{\phi e} m_e }{ a\,   m_\phi^2}\, \frac{\left(1 + b \, r^2 +c \, r^4 \log{ (r) } \right)}{\left(  1 + 4 \, r^2\right)^3} \,,\\
    C_{\rm loop-e} & = \frac{\lambda_{\chi\phi} \,  y^2_{\phi e} \left(  R \log
   \left(\frac{m_e^2}{m_\phi^2}\right)+ \left(1-2
   \frac{m_e^2}{m_\phi^2}\right) \log \left(\frac{m_\phi^2 (R+1)^2}{4
   m_e^2}\right)\right)}{8 \pi^{2} \, m_e R},
\end{align}
where $\lambda_{\chi\phi}\approx \alpha_D/\pi$ as in the previous section, $\alpha_{D} \equiv y^2_{\chi \phi} /4\pi$, and $R \equiv \sqrt{1 -4 m_e^2/m_\phi^2}$. In the limit of $m_\phi \rightarrow \infty$, the loop factor approaches $C_{\rm loop} \approx (\alpha_D/\pi) (y^2_{\phi e}/4\pi^{2}) \,m_e/m_\phi^2 $, as expected in the EFT limit. 

We note that the ``free parameters" $\lambda_{i}$ generate tree-level and one-loop contributions to the elastic scattering cross section as well. As noted earlier, for the purposes of our analysis, we focus on only the parameters relevant to Co-SIMP phenomenology $y_{\phi\chi}$ and $y_{\phi e}$, and consider how these parameters are constrained by experiments. 

In the leptophilic case, the halo population of DM has lower kinetic energy since it is lighter, and is therefore unable to trigger electron recoil signals in large noble gas detectors. However, several searches have sensitivity to the Co-SIMP parameter space today, such as the DM solar reflection analysis~\cite{Emken:2021lgc}. In the near future, promising new detector technologies will decrease the energy threshold for electronic recoils~\cite{Mei:2017etc, Hochberg:2017wce, Hochberg:2015fth} and be able to probe deeper into the parameter space.  We illustrate these constraints in Fig.~\ref{fig_PS_scalarLep}, and specifically display the spin-independent elastic scattering cross section predictions for thermal relic Co-SIMPs in Fig.~\ref{fig_targetsigma_scalarLep} to serve as benchmarks for future experiments.

\subsubsection{Monoenergetic Electron Recoil from Co-SIMP interaction}

In addition to elastic scattering, direct detection experiments may also potentially be sensitive to the monoenergetic recoil of electrons which occurs locally in the detector via the Co-SIMP process $\chi\chi e \to \chi^{\dagger} e$. Two ambient $\chi$ particles coincide with an electron in the detector and scatter, leading to an electron recoiling at 
\begin{equation}
    E_R = \frac{1}{2}\left(\frac{3m_{\chi}^{2}}{2m_{\chi} + m_{e}}\right) \simeq  0.3 \; \text{keV} \times \left( \frac{m_\chi}{ 10 \; \text{keV}}\right)^2  .
\end{equation}
While the event rate is expected to be somewhat lower, as it is quadratically sensitive to the DM halo density, monoenergetic signals typically enjoy a much lower background level and are easier to search for. The expected event rate at the site of a XENON1T-type experiment is approximately
    \begin{equation}
    \Gamma_{\rm 3\to 2} \simeq  \frac{2.6 \times 10^{-2}
}{\mathrm{yr}} \left( \frac{m_\chi}{ 10 \; \mathrm{keV}}\right)^{-2} \left( \frac{\langle \sigma v^2 \rangle}{ 10^{10} \; \mathrm{GeV}^{-5}} \right),
    \end{equation}
which is able to probe a substantial amount of our parameter space. We illustrate the present constraints from XENON1T, and the expected reach for XENONnT in Fig.~\ref{fig_PS_scalarLep}.

\subsubsection{Boosted Co-SIMPs from accumulation in the Earth}

Finally, another potential avenue to detect Co-SIMP dark matter is to search for signals from high energy DM produced locally via the Co-SIMP process $\chi\chi e \to \chi^{\dagger} e$ which then elastically scatters with the material of the detector via $\chi^{(\dagger)} e \to \chi^{(\dagger)} e$. This would induce quite different signatures in direct detection experiments in contrast to elastic scattering from non-relativistic ambient DM. 
    In particular, for Co-SIMP masses well below the electron mass, the boosted $\chi$ momenta are expected to be in the range of $p \approx \sqrt{3} m_\chi \left(1 - m_e/m_\chi \right)$, which would induce elastic scattering above threshold for a number of large volume experiments. 
    These signals, if detected, are likely to come from underground, as the bulk of local number-changing interactions occur while the DM is traveling within the Earth.  
    
    Just as the nucleophilic scenario, the event rate in this case then sensitively depends on the amount of DM accumulated in the Earth, which could be dramatically enhanced relative to the ambient halo abundance.  Repeating the basic calculations from the nucleophilic scenario, the expected event rate of boosted Co-SIMPs in the volume of the XENON1T experiment is given by,
\begin{align}
    \Gamma_{{\rm boosted} \; \chi e} =  \frac{1.35 }{\text{s}}  \, \frac{f_{\rm cap}}{4} \left(\frac{ v_\chi}{240 \, \text{km}/\text{s}}\right)  \left(\frac{ \rho_\chi \, \text{cm}^3 }{ m\chi} \right)  \left(\frac{ \sigma_{\chi e} }{ \text{pb}}\right)
     %\left(\frac{ M_T \, \text{GeV} }{ m_A \, \text{ton} }\right) 
     \left(\frac{ M_T }{ \text{ton} }\right) 
     \left(\frac{ \text{GeV} }{ m_A }\right) 
\end{align}
where the main difference relative to the nucleophilic case is that the elastic cross section $\sigma_{\chi e}$ has no coherent enhancement factor of $A^2$. This makes the expected sensitivity of this search less promising in the leptophillic scenario.

\subsection{Other Terrestrial Searches}

%\LX{Some preamble here.}

In this subsection, we discuss the constraints on our leptophilic scalar-mediated Co-SIMP model from other terrestrial experiments, including beam-dump experiments and the electron $g-2$ measurement. As in the nucleophilic scenario, independent of it being coupled to the DM, the existence of a light scalar $\phi$ that couples to the SM introduces phenomenology testable at precision or high energy particle physics experiments.

% The fact that the Co-SIMP interaction involves DM and SM particles in the interaction vertex which is crucial for setting its relic abundance, provides us with a unique opportunity to directly test this DM production mechanism in a laboratory experiment. The freezeout process provides us in this framework with target interaction rates for our searches.  In the following subsections, we will present a number of experimental opportunities and discuss their short-term optimization potential. 

%%%%%%%%%%%%%%%%%%%%%%%%%%%%%%%%%%%%%%%%%%%%%%%%%%%%%%%%%%%%%
\subsubsection{Beam Dump Experiments}

Since the models we consider include considerable couplings to the SM and light, stable, dark degrees of freedom, beam dump experiments are potentially a promising arena with which to discover Co-SIMP phenomenology. One potential avenue is via searching for missing energy signals from the production of dark matter $\chi\chi\chi$ from off-shell mediators $\phi$.  In the leptophilic Co-SIMP scenario however, we find the mean free path for inelastic (reverse) Co-SIMP production is too short to be detectable in current beam dump experiments; that is, the produced relativistic DM particles will likely recoil with the SM particles in a $\chi e \to \chi\chi e$ process before exiting the experiment, altering the observable signal. A more involved dedicated analysis would be needed to assess the discovery or constraining potential of experiments in this case.

However, there is another, simpler, possibility for testing this scenario in lepton based beam dump experiments~\cite{Marsicano:2018vin}: the production in this case is brem-emission of the scalar mediator $e + e \rightarrow e + e + \phi$, which, depending on the parameter region, can have a substantial lifetime, as its decay to photons is loop suppressed. The final decay process of $\phi \rightarrow \gamma \gamma$ can serve as a detection signature in the detector, located downstream beyond the electron or positron beam dump. We consider the sensitivity of the E137 experiment~\cite{Batell:2014mga}, which has $\sim 400 \text{ meter}$ decay length between the dump and the detector, and find that it can probe DM-SM couplings down to $y_{\phi e} \sim 10^{-2}$ in this Co-SIMP scenario.

%\ap{calculation to be done here}
%\js{ usually $e + e \rightarrow \phi \rightarrow  e + e$, but our $\phi$ is too light}
%\js{ $e + e \rightarrow \phi \rightarrow \gamma \gamma$, check if signal is detectable, and if decay length is testable. Only model which seems testable with beam dump.. }

%%%%%%%%%%%%%%%%%%%%%%%%%%%%%%%%%%%%%%%%%%%%%%%%%%%%%%%%%%%%%
\subsubsection{Electron $g-2$}

The anomalous magnetic dipole moment of the electron has been measured extremely precisely, with an experimental value of $a_{e} = 0.00115965218073(28)$~\cite{Hanneke:2010au}. This measurement in turn strongly constrains any new physics that couples to the electron and contributes to the electron $g-2$. In our model, this role is fulfilled by the scalar mediator $\phi$. For a scalar, the $g - 2$ contribution is given by
\begin{equation}
    F_{2}(0) = \frac{y_{\phi e}^{2}}{8\pi^{2}}\int_{0}^{1}dz\frac{(1+z)(1-z)^{2}}{(1-z)^{2} + z\xi^{2}},
\end{equation}
where we have defined $\xi \equiv m_\phi/m_e$.
Integrating this, we find
\begin{equation}
\begin{split}
    &g - 2 = \frac{y_{\phi e}^{2}}{4\pi^{2}}\Bigg[\frac{1}{\sqrt{\xi^{2} - 4}}\bigg(\xi(\xi^{4} - 5\xi^{2} + 4)\text{tanh}^{-1}\bigg(\frac{\xi}{\sqrt{\xi^{2}-4}}\bigg)\bigg) \\
    &+ \frac{1}{2}\bigg(3 -2\xi^{2} +(\xi^{2} - 3)\xi^{2}\text{log}\xi^{2} - 2\sqrt{\xi^{2} - 4}(\xi^{2} - 1)\xi\text{tanh}^{-1}\bigg(\frac{\xi^{2} - 2}{\xi\sqrt{\xi^{2} - 4}}\bigg)  \bigg)\Bigg].
\end{split}
\end{equation}
Since we are interested in light scalars, we can expand our result for small $\xi$. Doing so, we obtain
\begin{equation}
    g - 2 \approx \frac{y_{\phi e}^{2}}{4\pi^{2}}\bigg(\frac{3}{2} - \pi\xi\bigg).
\end{equation}

For light scalars, the $g-2$ contribution is constant with a subleading linear dependence on the mass ratio $\xi$, which we see in Fig.~\ref{fig_PS_scalarLep}. The scalar can also couple to muons in the case of universal leptophilic couplings, which would lead to an analogous contribution to the muon magnetic dipole moment, but we find this much less constraining for the parameter space of interest~\cite{Abi:2021gix}.

%%%%%%%%%%%%%%%%%%%%%%%%%%%%%%%%%%%%%%%%%%%%%%%%%%%%%%%%%%%%%

\subsection{Parameter Space}
\label{sec:scalar-lepto-param}
\begin{figure}[h!]
    \caption{     Parameter space of our leptophilic Co-SIMP model with a scalar mediator. Current constraints~\cite{Hanneke:2010au,Emken:2019tni,XENON:2019gfn,XENONCollaborationSS:2021sgk} are shown as the \textit{shaded regions} and the projected sensitivities (XENONnT~\cite{XENONCollaboration:2022kmb} and superconducting aluminium~\cite{Hochberg:2015fth}) as \textit{dashed lines}. The \textit{dark red lines} corresponds to the parameter space that gives the correct relic abundance that matches observations, assuming different values for $\mu_R \equiv m_\phi/m_\chi$. 
    For this scenario, stellar cooling~\cite{Grifols:1986fc,1989MPLA....4..311G} places a stronger constraint than fifth force searches~\cite{Murata:2014nra}. In particular, fifth-force searches constrain mediators that are lighter than the ones we consider here. As indicated by the \textit{dashed magenta line}, the elastic scattering search for boosted Co-SIMPs is not competitive in this scenario. 
    }
    \resizebox{0.6\columnwidth}{!}{
    \includegraphics[scale=0.45]{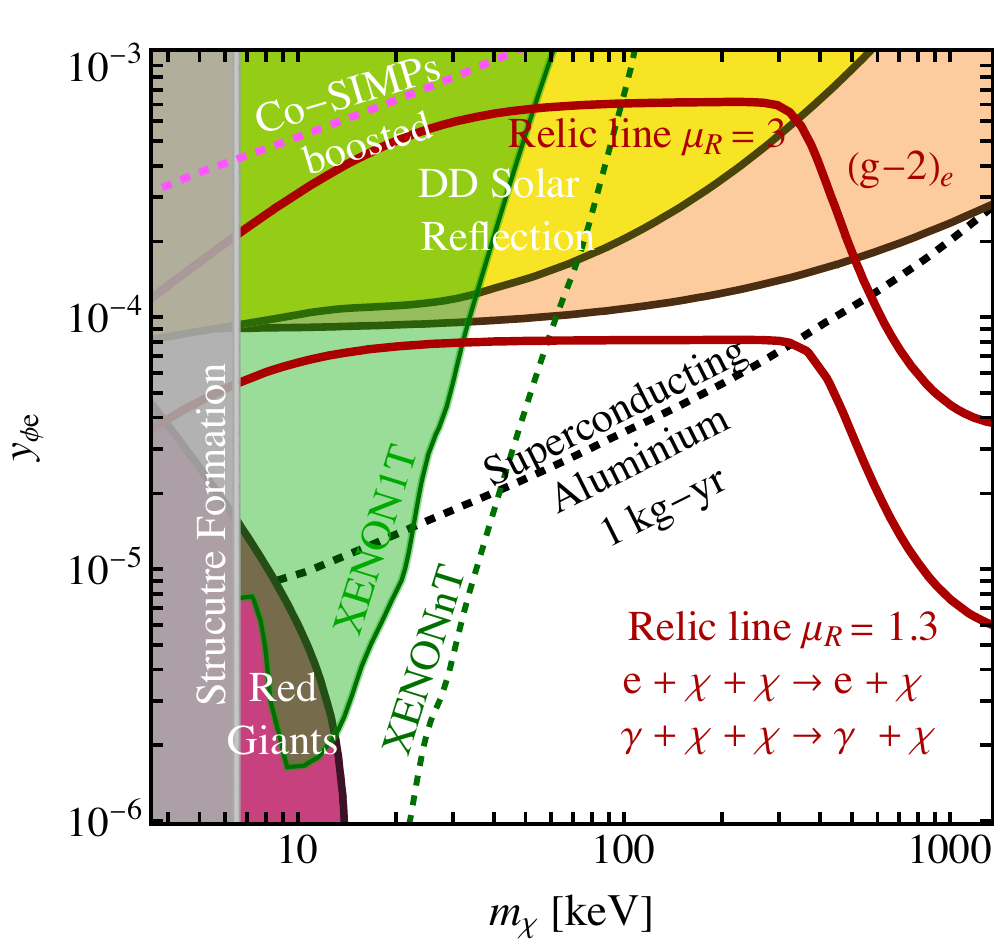}
       }
     \label{fig_PS_scalarLep}
\end{figure}
\begin{figure}[h!]
    \caption{
    Similar to Fig.~\ref{fig:nucleosigmael} but for the leptophilic case, showcasing the reach of future direct detection experiments in the Co-SIMP parameter space~\cite{Hochberg:2015fth}. Currently, the solar reflection limits~\cite{An:2021qdl} seem to provide the best sensitivity to the elastic scattering cross section and are shown in the \textit{yellow contour}. 
    The \textit{dark red lines} show the benchmark cross sections in the Co-SIMP scenario such that we recover the correct present-day relic abundance with various choices of $\mu_R\equiv m_\phi/m_\chi$. 
    }
    \resizebox{0.6\columnwidth}{!}{
    \includegraphics[scale=0.45]{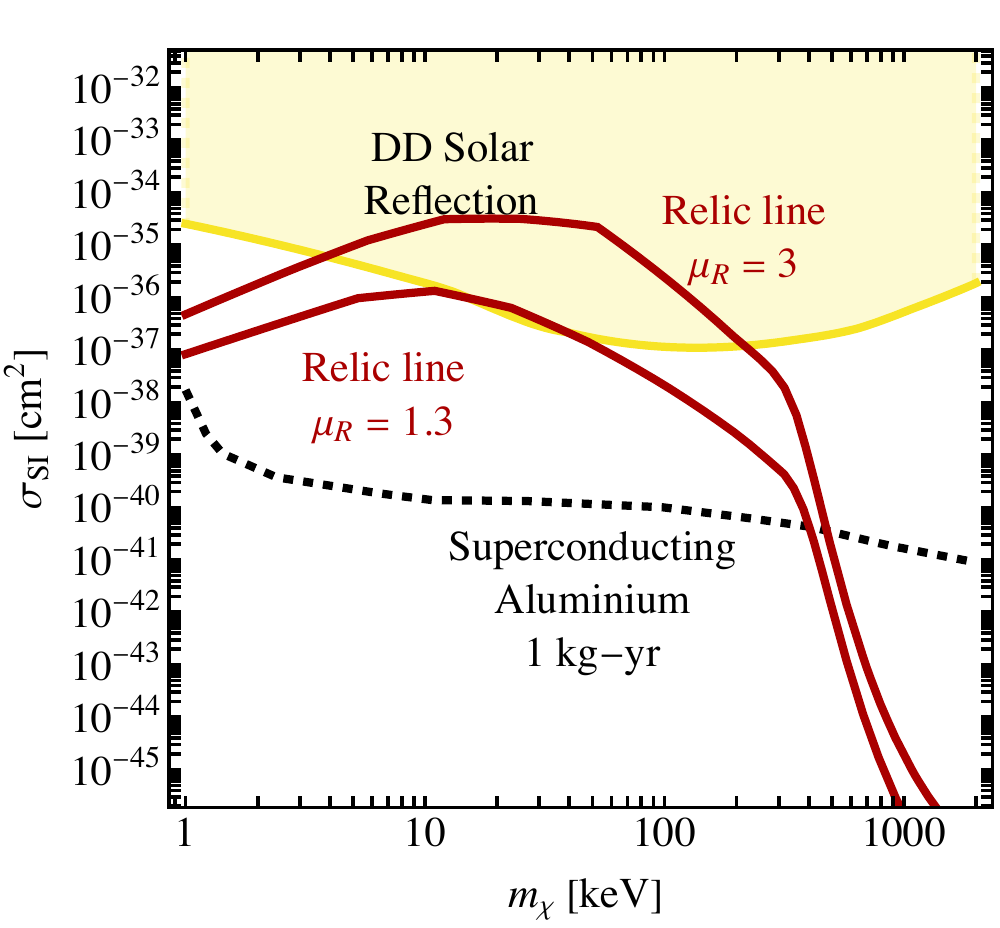}
       }
     \label{fig_targetsigma_scalarLep}
\end{figure}

In Fig.~\ref{fig_PS_scalarLep} we show the parameter space of our scalar-mediated leptophilic Co-SIMP model including constraints from presently available experiments (solid) and projected sensitivities of future experimental searches (dashed). The predictions for a leptophilic Co-SIMP with a thermal history constituting 100\% of the DM abundance are shown in red for various choices of the dark sector mass hierarchy $\mu_R \equiv m_\phi/m_\chi$.  As in the nucleophilic discussion, all constraints are presented in the $m_\chi$-$y_{\phi e}$ space and assume a benchmark mass ratio $\mu_R = 1.3$.

We find that the parameter space is unconstrained in the region of large masses and small couplings.  The combination of the constraints from structure formation and red giant cooling~\cite{Grifols:1986fc,1989MPLA....4..311G}, as well as null observations of monoenergetic electron recoils observed at  XENON1T~\cite{XENONCollaborationSS:2021sgk} excludes the region of $m_{\chi}$ below few tens of keV.   The electron $g-2$ measurement~\cite{Hanneke:2010au} excludes the region with $y_{\phi e} \gtrsim 10^{-4}$, subsuming present-day constraints on elastic scattering from solar reflection analyses~\cite{Emken:2021lgc}.   Future XENONnT monoenergetic recoil searches~\cite{XENONCollaboration:2022kmb} could restrict the open parameter space to even higher masses, whereas the elastic scattering signatures can become within reach for new low-threshold experiments~\cite{Hochberg:2015fth,Hochberg:2017wce,Mei:2017etc}. As an example, we show the expected sensitivity of a 1 kg-yr detector based on superconducting aluminium technology~\cite{Hochberg:2015fth}. Finally, as discussed above, the Co-SIMP process inside the Earth can lead to a flux of boosted particles, with elastic scattering above the detection threshold of current liquid noble gas detectors. However, in the leptophilic case, the currently achievable sensitivity, which is shown as the dashed magenta line  in Fig.~\ref{fig_PS_scalarLep}, is not competitive.  

In aggregate, the thermal prediction for the leptophilic scenario is much more well-constrained than the nucleophilic case, with most of the parameter space for $m_\chi \lesssim 500$ keV ruled out except for models with quasi-degenreate masses, $\mu_R \approx 1$.  Models with heavier ($\sim$ MeV) DM require smaller couplings to achieve the correct abundance, and are largely unconstrained by present limits, but are precisely within the sensitivity targets of future low-threshold technologies.

As in the nucleophilic case, it is worth remarking upon that almost the entirety of the thermal relic prediction can be tested with next-generation direct detection technology with a combination of searches for inelastic and elastic, and low-  and high-threshold signatures.  In Fig.~\ref{fig_targetsigma_scalarLep} we show the future direct detection landscape along specifically the axis of the elastic scattering cross section and demonstrate where the thermal relic expectation lies therein. As shown, the expected scattering strengths, while largely inaccessible by current experiments~\cite{Emken:2019tni}, represent an interesting target for future experimental efforts~\cite{Mei:2017etc,Hochberg:2017wce,Hochberg:2015fth}, and the potential for both elastic and inelastic detection signals as well as low- and high- threshold elastic scattering give this DM candidate a set of very unique and predictive signatures and make it an interesting case study in the direct detection arena.

\section{Conclusions}
\label{sec:conclusions}

The Co-SIMPs framework ($\chi + \chi + \text{SM} \rightarrow \chi + \text{SM}$ depicted in Fig.~\ref{fig:cosimp}) is a novel thermal-production mechanism for dark matter~\cite{Smirnov:2020zwf}. 
Like other thermal DM scenarios, the rate of the 3-to-2 process is determined by the dark matter freeze-out condition and can directly be tested in direct detection experiments. However, the unique topology of the interactions involved allow for vastly different relationships between the expected depletion and scattering rates compared to the standard WIMP scenario, and naturally predict lower mass candidates. Additionally, in contrast to the SIMP scenario, the dark and visible sectors are kept in kinetic equilibrium in this scenario, preventing an overheating of the dark sector through conversion of rest mass into kinetic energy.  

In this work we, for the first time, study UV-complete models for the Co-SIMP framework, where the 3-to-2 process (Fig.~\ref{fig:cosimp}) is mediated by a scalar mediator. We study two cases: a mediator that couples to nucleons (Sec.~\ref{sec:scalar_nucleo}) and a mediator that couples to electrons (Sec.~\ref{sec:scalar_lepto}). For each case, we study its freeze-out in detail and possible constraints from cosmology, astrophysics, and terrestrial experiments extensively.

With these scalar-mediated scenarios, we study the simplest class of models to realize the Co-SIMP process which has a dark sector with two particles: dark matter $\chi$ and the scalar mediator $\phi$ that couples to SM particles. The $\chi$ has a charge 1 and $\phi$ is neutral under the $\mathbb{Z}_{3}$ symmetry, which protects the stability of $\chi$. 
In the nucleophilic scenario, the topology of the Co-SIMP interaction is realized in a $\chi \chi N \rightarrow \chi^\dagger N$ process below the QCD scale.  The Co-SIMP-nucleon coupling is driven by the interaction with the up and down quarks. The typical mass ranges for $\chi$ and $\phi$ are $\sim$ MeV--GeV. The parameter space (Fig.~\ref{fig:nucleoparameterspace}) is mostly unconstrained for $m_{\chi} > 0.01$~GeV in terms of current experimental results. 
To further search for Co-SIMPs in this scenario, a detector based on superfluid helium~\cite{Ruppin:2014bra, Schutz:2016tid, Hertel:2018aal} would test a significant fraction of the open parameter space. Moreover, a dedicated beam-dump search for inelastic production of Co-SIMPs could probe the region of larger couplings and larger masses ($m_{\chi} \sim 1$~GeV).

In the leptophilic scenario, the typical mass ranges for $\chi$ and $\phi$ are $\lesssim$ MeV. The parameter space (Fig.~\ref{fig_PS_scalarLep}) is more constrained than the nucleophilic case. The combination of the constraints from structure formation, red giant cooling, and XENON1T direct detection excludes the region of $m_{\chi}$ below few tens of keV. The electron $g-2$ excludes the region with $y_{\phi e} \gtrsim 10^{-4}$. However, for $m_{\chi} \gtrsim 10$~keV and $y_{\phi e} \lesssim 10^{-4}$, the parameter space is mostly open. Future XENONnT results will further constrain the open parameter space and push us to higher masses. However, this scenario is pertinently interesting in light of dramatic technological developments in low-threshold direct detection, and much of the remaining parameter space may be probed in a search with next-generation technologies once they mature. 

This paper focuses on the scenario with a scalar mediator. This scenario is compelling as it directly realizes the effective diagram topology in Fig.~\ref{fig:cosimp}, but runs into theoretical issues inherent to the generation and stabilization of light scalar masses. In upcoming work~\cite{Co-SIMP_Vector}, we will consider a separate completion involving fermionic Co-SIMPs and a vector mediator, which alleviates many of the model-building issues associated with light scalars. The vector scenario leads to new experimental and observational signatures which we will examine in more detail. The models presented here and in our upcoming work are the first attempts at UV-completing the Co-SIMP mechanism, which is one facet in a broader vein of work generalizing the landscape of thermal DM beyond the WIMP scenario. Other UV-complete models may also exist, and a careful exploration of the space of models will be important to quantify how low the thermal dark matter mass can reach within this framework.

\section*{Acknowledgments}

We thank Arindam Bhattacharya, Rouven Essig, Gordan Krnjaic, Matthew Reece, Tracy Slatyer, Xiaoyuan Zhang, Jure Zupan, and especially Stephen West and John Beacom for helpful discussions.
AP is supported in part by the DOE Grant DE-SC0013607, the Alfred P. Sloan Foundation Grant No. G-2019-12504, the NSF Grant PHY2210533, and the Simons Foundation Grant No. 623940.
JS was largely supported by a Feodor Lynen Fellowship from the Alexander von Humboldt foundation. JS thanks The Ohio State University (OSU), the Center for Cosmology and Astroparticle Physics (CCAPP) and Stockholm University for hospitality, and support during large parts of this project.  
WLX is supported by the U.S. Department of Energy under
Contract DE-AC02-05CH11231. 
BZ is supported by the Simons Foundation.

% \appendix
% \section{A}
% \label{appx:thermo}

\afterpage{\clearpage}
\bibliographystyle{utphys_modified}
\bibliography{references}

\providecommand{\href}[2]{#2}\begingroup\raggedright\begin{thebibliography}{100}

\bibitem{Steigman:1984ac}
G.~Steigman and M.~S. Turner, ``{Cosmological Constraints on the Properties of
  Weakly Interacting Massive Particles},''
  \href{http://dx.doi.org/10.1016/0550-3213(85)90537-1}{{\em Nucl. Phys. B}
  {\bfseries 253} (1985) 375--386}.

\bibitem{Jungman:1995df}
G.~Jungman, M.~Kamionkowski, and K.~Griest, ``{Supersymmetric dark matter},''
  \href{http://dx.doi.org/10.1016/0370-1573(95)00058-5}{{\em Phys. Rept.}
  {\bfseries 267} (1996) 195--373},
  \href{http://arxiv.org/abs/hep-ph/9506380}{{\ttfamily arXiv:hep-ph/9506380}}.

\bibitem{Bertone:2004pz}
G.~Bertone, D.~Hooper, and J.~Silk, ``{Particle dark matter: Evidence,
  candidates and constraints},''
  \href{http://dx.doi.org/10.1016/j.physrep.2004.08.031}{{\em Phys. Rept.}
  {\bfseries 405} (2005) 279--390},
  \href{http://arxiv.org/abs/hep-ph/0404175}{{\ttfamily arXiv:hep-ph/0404175}}.

\bibitem{Bertone:2016nfn}
G.~Bertone and D.~Hooper, ``{History of dark matter},''
  \href{http://dx.doi.org/10.1103/RevModPhys.90.045002}{{\em Rev. Mod. Phys.}
  {\bfseries 90} no.~4, (2018) 045002},
  \href{http://arxiv.org/abs/1605.04909}{{\ttfamily arXiv:1605.04909
  [astro-ph.CO]}}.

\bibitem{DeLuca:2018mzn}
V.~De~Luca, A.~Mitridate, M.~Redi, J.~Smirnov, and A.~Strumia, ``{Colored Dark
  Matter},'' \href{http://dx.doi.org/10.1103/PhysRevD.97.115024}{{\em Phys.
  Rev. D} {\bfseries 97} no.~11, (2018) 115024},
  \href{http://arxiv.org/abs/1801.01135}{{\ttfamily arXiv:1801.01135
  [hep-ph]}}.

\bibitem{Gross:2018zha}
C.~Gross, A.~Mitridate, M.~Redi, J.~Smirnov, and A.~Strumia, ``{Cosmological
  Abundance of Colored Relics},''
  \href{http://dx.doi.org/10.1103/PhysRevD.99.016024}{{\em Phys. Rev. D}
  {\bfseries 99} no.~1, (2019) 016024},
  \href{http://arxiv.org/abs/1811.08418}{{\ttfamily arXiv:1811.08418
  [hep-ph]}}.

\bibitem{Harz:2018csl}
J.~Harz and K.~Petraki, ``{Radiative bound-state formation in unbroken
  perturbative non-Abelian theories and implications for dark matter},''
  \href{http://dx.doi.org/10.1007/JHEP07(2018)096}{{\em JHEP} {\bfseries 07}
  (2018) 096}, \href{http://arxiv.org/abs/1805.01200}{{\ttfamily
  arXiv:1805.01200 [hep-ph]}}.

\bibitem{Griest:1989wd}
K.~Griest and M.~Kamionkowski, ``{Unitarity Limits on the Mass and Radius of
  Dark Matter Particles},''
  \href{http://dx.doi.org/10.1103/PhysRevLett.64.615}{{\em Phys. Rev. Lett.}
  {\bfseries 64} (1990) 615}.

\bibitem{Smirnov:2019ngs}
J.~Smirnov and J.~F. Beacom, ``{TeV-Scale Thermal WIMPs: Unitarity and its
  Consequences},'' \href{http://dx.doi.org/10.1103/PhysRevD.100.043029}{{\em
  Phys. Rev. D} {\bfseries 100} no.~4, (2019) 043029},
  \href{http://arxiv.org/abs/1904.11503}{{\ttfamily arXiv:1904.11503
  [hep-ph]}}.

\bibitem{Asadi:2021pwo}
P.~Asadi, E.~D. Kramer, E.~Kuflik, G.~W. Ridgway, T.~R. Slatyer, and
  J.~Smirnov, ``{Thermal squeezeout of dark matter},''
  \href{http://dx.doi.org/10.1103/PhysRevD.104.095013}{{\em Phys. Rev. D}
  {\bfseries 104} no.~9, (2021) 095013},
  \href{http://arxiv.org/abs/2103.09827}{{\ttfamily arXiv:2103.09827
  [hep-ph]}}.

\bibitem{Asadi:2021yml}
P.~Asadi, E.~D. Kramer, E.~Kuflik, G.~W. Ridgway, T.~R. Slatyer, and
  J.~Smirnov, ``{Accidentally Asymmetric Dark Matter},''
  \href{http://dx.doi.org/10.1103/PhysRevLett.127.211101}{{\em Phys. Rev.
  Lett.} {\bfseries 127} no.~21, (2021) 211101},
  \href{http://arxiv.org/abs/2103.09822}{{\ttfamily arXiv:2103.09822
  [hep-ph]}}.

\bibitem{Asadi:2021bxp}
P.~Asadi, T.~R. Slatyer, and J.~Smirnov, ``{WIMPs without weakness: Generalized
  mass window with entropy injection},''
  \href{http://dx.doi.org/10.1103/PhysRevD.106.015012}{{\em Phys. Rev. D}
  {\bfseries 106} no.~1, (2022) 015012},
  \href{http://arxiv.org/abs/2111.11444}{{\ttfamily arXiv:2111.11444
  [hep-ph]}}.

\bibitem{Asadi:2022vkc}
P.~Asadi, E.~D. Kramer, E.~Kuflik, T.~R. Slatyer, and J.~Smirnov, ``{Glueballs
  in a thermal squeezeout model},''
  \href{http://dx.doi.org/10.1007/JHEP07(2022)006}{{\em JHEP} {\bfseries 07}
  (2022) 006}, \href{http://arxiv.org/abs/2203.15813}{{\ttfamily
  arXiv:2203.15813 [hep-ph]}}.

\bibitem{Smirnov:2022tcg}
J.~Smirnov, \href{http://dx.doi.org/10.21468/SciPostPhysProc}{``{Dark Matter
  Bound States: A Window into the Early Universe},''} in {\em {14th
  International Workshop on the Identification of Dark Matter 2022}}.
\newblock 12, 2022.
\newblock \href{http://arxiv.org/abs/2212.14361}{{\ttfamily arXiv:2212.14361
  [hep-ph]}}.

\bibitem{Asadi:2022njl}
P.~Asadi {\em et~al.}, ``{Early-Universe Model Building},''
  \href{http://arxiv.org/abs/2203.06680}{{\ttfamily arXiv:2203.06680
  [hep-ph]}}.

\bibitem{Smirnov:2022zip}
J.~Smirnov, A.~Goobar, T.~Linden, and E.~M\"ortsell, ``{White Dwarfs in Dwarf
  Spheroidal Galaxies: A New Class of Compact-Dark-Matter Detectors},''
  \href{http://arxiv.org/abs/2211.00013}{{\ttfamily arXiv:2211.00013
  [astro-ph.CO]}}.

\bibitem{DAgnolo:2017dbv}
R.~T. D'Agnolo, D.~Pappadopulo, and J.~T. Ruderman, ``{Fourth Exception in the
  Calculation of Relic Abundances},''
  \href{http://dx.doi.org/10.1103/PhysRevLett.119.061102}{{\em Phys. Rev.
  Lett.} {\bfseries 119} no.~6, (2017) 061102},
  \href{http://arxiv.org/abs/1705.08450}{{\ttfamily arXiv:1705.08450
  [hep-ph]}}.

\bibitem{DAgnolo:2019zkf}
R.~T. D'Agnolo, D.~Pappadopulo, J.~T. Ruderman, and P.-J. Wang, ``{Thermal
  Relic Targets with Exponentially Small Couplings},''
  \href{http://dx.doi.org/10.1103/PhysRevLett.124.151801}{{\em Phys. Rev.
  Lett.} {\bfseries 124} no.~15, (2020) 151801},
  \href{http://arxiv.org/abs/1906.09269}{{\ttfamily arXiv:1906.09269
  [hep-ph]}}.

\bibitem{Hochberg:2014dra}
Y.~Hochberg, E.~Kuflik, T.~Volansky, and J.~G. Wacker, ``{Mechanism for Thermal
  Relic Dark Matter of Strongly Interacting Massive Particles},''
  \href{http://dx.doi.org/10.1103/PhysRevLett.113.171301}{{\em Phys. Rev.
  Lett.} {\bfseries 113} (2014) 171301},
  \href{http://arxiv.org/abs/1402.5143}{{\ttfamily arXiv:1402.5143 [hep-ph]}}.

\bibitem{Cline:2017tka}
J.~M. Cline, H.~Liu, T.~Slatyer, and W.~Xue, ``{Enabling Forbidden Dark
  Matter},'' \href{http://dx.doi.org/10.1103/PhysRevD.96.083521}{{\em Phys.
  Rev. D} {\bfseries 96} no.~8, (2017) 083521},
  \href{http://arxiv.org/abs/1702.07716}{{\ttfamily arXiv:1702.07716
  [hep-ph]}}.

\bibitem{Kopp:2016yji}
J.~Kopp, J.~Liu, T.~R. Slatyer, X.-P. Wang, and W.~Xue, ``{Impeded Dark
  Matter},'' \href{http://dx.doi.org/10.1007/JHEP12(2016)033}{{\em JHEP}
  {\bfseries 12} (2016) 033}, \href{http://arxiv.org/abs/1609.02147}{{\ttfamily
  arXiv:1609.02147 [hep-ph]}}.

\bibitem{Farina:2016llk}
M.~Farina, D.~Pappadopulo, J.~T. Ruderman, and G.~Trevisan, ``{Phases of
  Cannibal Dark Matter},''
  \href{http://dx.doi.org/10.1007/JHEP12(2016)039}{{\em JHEP} {\bfseries 12}
  (2016) 039}, \href{http://arxiv.org/abs/1607.03108}{{\ttfamily
  arXiv:1607.03108 [hep-ph]}}.

\bibitem{Dey:2016qgf}
U.~K. Dey, T.~N. Maity, and T.~S. Ray, ``{Light Dark Matter through Assisted
  Annihilation},'' \href{http://dx.doi.org/10.1088/1475-7516/2017/03/045}{{\em
  JCAP} {\bfseries 03} (2017) 045},
  \href{http://arxiv.org/abs/1612.09074}{{\ttfamily arXiv:1612.09074
  [hep-ph]}}.

\bibitem{DAgnolo:2018wcn}
R.~T. D'Agnolo, C.~Mondino, J.~T. Ruderman, and P.-J. Wang, ``{Exponentially
  Light Dark Matter from Coannihilation},''
  \href{http://dx.doi.org/10.1007/JHEP08(2018)079}{{\em JHEP} {\bfseries 08}
  (2018) 079}, \href{http://arxiv.org/abs/1803.02901}{{\ttfamily
  arXiv:1803.02901 [hep-ph]}}.

\bibitem{Kim:2019udq}
H.~Kim and E.~Kuflik, ``{Superheavy Thermal Dark Matter},''
  \href{http://dx.doi.org/10.1103/PhysRevLett.123.191801}{{\em Phys. Rev.
  Lett.} {\bfseries 123} no.~19, (2019) 191801},
  \href{http://arxiv.org/abs/1906.00981}{{\ttfamily arXiv:1906.00981
  [hep-ph]}}.

\bibitem{Ghosh:2023ocl}
D.~K. Ghosh, P.~Ghosh, and S.~Jeesun, ``{CMB signature of non-thermal Dark
  Matter produced from self-interacting dark sector},''
  \href{http://arxiv.org/abs/2301.13754}{{\ttfamily arXiv:2301.13754
  [hep-ph]}}.

\bibitem{Smirnov:2020zwf}
J.~Smirnov and J.~F. Beacom, ``{New Freezeout Mechanism for Strongly
  Interacting Dark Matter},''
  \href{http://dx.doi.org/10.1103/PhysRevLett.125.131301}{{\em Phys. Rev.
  Lett.} {\bfseries 125} no.~13, (2020) 131301},
  \href{http://arxiv.org/abs/2002.04038}{{\ttfamily arXiv:2002.04038
  [hep-ph]}}.

\bibitem{Essig:2022dfa}
R.~Essig, G.~K. Giovanetti, N.~Kurinsky, D.~McKinsey, K.~Ramanathan,
  K.~Stifter, and T.-T. Yu, ``{Snowmass2021 Cosmic Frontier: The landscape of
  low-threshold dark matter direct detection in the next decade},'' in {\em
  {2022 Snowmass Summer Study}}.
\newblock 3, 2022.
\newblock \href{http://arxiv.org/abs/2203.08297}{{\ttfamily arXiv:2203.08297
  [hep-ph]}}.

\bibitem{Co-SIMP_Vector}
A.~Parikh, J.~Smirnov, W.~L. Xu, and B.~Zhou, {\em {Fermionic Co-SIMP Dark
  Matter: Models and Sensitivities (To Appear)}}.

\bibitem{Harlow:2022gzl}
D.~Harlow, B.~Heidenreich, M.~Reece, and T.~Rudelius, ``{The Weak Gravity
  Conjecture: A Review},'' \href{http://arxiv.org/abs/2201.08380}{{\ttfamily
  arXiv:2201.08380 [hep-th]}}.

\bibitem{Baek:2013qwa}
S.~Baek, P.~Ko, and W.-I. Park, ``{Singlet Portal Extensions of the Standard
  Seesaw Models to a Dark Sector with Local Dark Symmetry},''
  \href{http://dx.doi.org/10.1007/JHEP07(2013)013}{{\em JHEP} {\bfseries 07}
  (2013) 013}, \href{http://arxiv.org/abs/1303.4280}{{\ttfamily arXiv:1303.4280
  [hep-ph]}}.

\bibitem{Ko:2014nha}
P.~Ko and Y.~Tang, ``{Self-interacting scalar dark matter with local $Z_3$
  symmetry},'' \href{http://dx.doi.org/10.1088/1475-7516/2014/05/047}{{\em
  JCAP} {\bfseries 05} (2014) 047},
  \href{http://arxiv.org/abs/1402.6449}{{\ttfamily arXiv:1402.6449 [hep-ph]}}.

\bibitem{Frigerio:2022kyu}
M.~Frigerio, N.~Grimbaum-Yamamoto, and T.~Hambye, ``{Dark matter from the
  centre of SU(N)},'' \href{http://arxiv.org/abs/2212.11918}{{\ttfamily
  arXiv:2212.11918 [hep-ph]}}.

\bibitem{Cline:2013gha}
J.~M. Cline, K.~Kainulainen, P.~Scott, and C.~Weniger, ``{Update on scalar
  singlet dark matter},''
  \href{http://dx.doi.org/10.1103/PhysRevD.88.055025}{{\em Phys. Rev. D}
  {\bfseries 88} (2013) 055025},
  \href{http://arxiv.org/abs/1306.4710}{{\ttfamily arXiv:1306.4710 [hep-ph]}}.
  [Erratum: Phys.Rev.D 92, 039906 (2015)].

\bibitem{Shifman:1978zn}
M.~A. Shifman, A.~I. Vainshtein, and V.~I. Zakharov, ``{Remarks on Higgs Boson
  Interactions with Nucleons},''
  \href{http://dx.doi.org/10.1016/0370-2693(78)90481-1}{{\em Phys. Lett. B}
  {\bfseries 78} (1978) 443--446}.

\bibitem{Elor:2021swj}
G.~Elor, R.~McGehee, and A.~Pierce, ``{Maximizing Direct Detection with HYPER
  Dark Matter},'' \href{http://arxiv.org/abs/2112.03920}{{\ttfamily
  arXiv:2112.03920 [hep-ph]}}.

\bibitem{Mandal:2022yym}
S.~Mandal and N.~Sehgal, ``{Mass-varying Dark Matter from a Phase
  Transition},'' \href{http://arxiv.org/abs/2212.07884}{{\ttfamily
  arXiv:2212.07884 [hep-ph]}}.

\bibitem{Coleman:1977py}
S.~R. Coleman, ``{The Fate of the False Vacuum. 1. Semiclassical Theory},''
  \href{http://dx.doi.org/10.1103/PhysRevD.16.1248}{{\em Phys. Rev. D}
  {\bfseries 15} (1977) 2929--2936}. [Erratum: Phys.Rev.D 16, 1248 (1977)].

\bibitem{Callan:1977pt}
C.~G. Callan, Jr. and S.~R. Coleman, ``{The Fate of the False Vacuum. 2. First
  Quantum Corrections},''
  \href{http://dx.doi.org/10.1103/PhysRevD.16.1762}{{\em Phys. Rev. D}
  {\bfseries 16} (1977) 1762--1768}.

\bibitem{Coleman:1980aw}
S.~R. Coleman and F.~De~Luccia, ``{Gravitational Effects on and of Vacuum
  Decay},'' \href{http://dx.doi.org/10.1103/PhysRevD.21.3305}{{\em Phys. Rev.
  D} {\bfseries 21} (1980) 3305}.

\bibitem{Hall:2009bx}
L.~J. Hall, K.~Jedamzik, J.~March-Russell, and S.~M. West, ``{Freeze-In
  Production of FIMP Dark Matter},''
  \href{http://dx.doi.org/10.1007/JHEP03(2010)080}{{\em JHEP} {\bfseries 03}
  (2010) 080}, \href{http://arxiv.org/abs/0911.1120}{{\ttfamily arXiv:0911.1120
  [hep-ph]}}.

\bibitem{Sachdev:2020bkk}
S.~Sachdev, T.~Regimbau, and B.~S. Sathyaprakash, ``{Subtracting compact binary
  foreground sources to reveal primordial gravitational-wave backgrounds},''
  \href{http://dx.doi.org/10.1103/PhysRevD.102.024051}{{\em Phys. Rev. D}
  {\bfseries 102} no.~2, (2020) 024051},
  \href{http://arxiv.org/abs/2002.05365}{{\ttfamily arXiv:2002.05365 [gr-qc]}}.

\bibitem{Sharma:2020btq}
A.~Sharma and J.~Harms, ``{Searching for cosmological gravitational-wave
  backgrounds with third-generation detectors in the presence of an
  astrophysical foreground},''
  \href{http://dx.doi.org/10.1103/PhysRevD.102.063009}{{\em Phys. Rev. D}
  {\bfseries 102} no.~6, (2020) 063009},
  \href{http://arxiv.org/abs/2006.16116}{{\ttfamily arXiv:2006.16116 [gr-qc]}}.

\bibitem{Biscoveanu:2020gds}
S.~Biscoveanu, C.~Talbot, E.~Thrane, and R.~Smith, ``{Measuring the primordial
  gravitational-wave background in the presence of astrophysical
  foregrounds},'' \href{http://dx.doi.org/10.1103/PhysRevLett.125.241101}{{\em
  Phys. Rev. Lett.} {\bfseries 125} (2020) 241101},
  \href{http://arxiv.org/abs/2009.04418}{{\ttfamily arXiv:2009.04418
  [astro-ph.HE]}}.

\bibitem{Zhou:2022otw}
B.~Zhou, L.~Reali, E.~Berti, M.~\c{C}al\i{}\c{s}kan, C.~Creque-Sarbinowski,
  M.~Kamionkowski, and B.~S. Sathyaprakash, ``{Compact Binary Foreground
  Subtraction in Next-Generation Ground-Based Observatories},''
  \href{http://arxiv.org/abs/2209.01221}{{\ttfamily arXiv:2209.01221 [gr-qc]}}.

\bibitem{Zhou:2022nmt}
B.~Zhou, L.~Reali, E.~Berti, M.~\c{C}al\i{}\c{s}kan, C.~Creque-Sarbinowski,
  M.~Kamionkowski, and B.~S. Sathyaprakash, ``{Subtracting Compact Binary
  Foregrounds to Search for Subdominant Gravitational-Wave Backgrounds in
  Next-Generation Ground-Based Observatories},''
  \href{http://arxiv.org/abs/2209.01310}{{\ttfamily arXiv:2209.01310 [gr-qc]}}.

\bibitem{Zhong:2022ylh}
H.~Zhong, R.~Ormiston, and V.~Mandic, ``{Detecting cosmological gravitational
  waves background after removal of compact binary coalescences in future
  gravitational wave detectors},''
  \href{http://arxiv.org/abs/2209.11877}{{\ttfamily arXiv:2209.11877 [gr-qc]}}.

\bibitem{Racco:2022bwj}
D.~Racco and D.~Poletti, ``{Precision cosmology with primordial GW backgrounds
  in presence of astrophysical foregrounds},''
  \href{http://arxiv.org/abs/2212.06602}{{\ttfamily arXiv:2212.06602
  [astro-ph.CO]}}.

\bibitem{Pan:2023naq}
Z.~Pan and H.~Yang, ``{Detecting Primordial Stochastic Gravitational Waves with
  Reduced Astrophysical Foregrounds},''
  \href{http://arxiv.org/abs/2301.04529}{{\ttfamily arXiv:2301.04529 [gr-qc]}}.

\bibitem{Hsyu:2020uqb}
T.~Hsyu, R.~J. Cooke, J.~X. Prochaska, and M.~Bolte, ``{The PHLEK Survey: A New
  Determination of the Primordial Helium Abundance},''
  \href{http://dx.doi.org/10.3847/1538-4357/ab91af}{{\em Astrophys. J.}
  {\bfseries 896} no.~1, (2020) 77},
  \href{http://arxiv.org/abs/2005.12290}{{\ttfamily arXiv:2005.12290
  [astro-ph.GA]}}.

\bibitem{Berlin:2019pbq}
A.~Berlin, N.~Blinov, and S.~W. Li, ``{Dark Sector Equilibration During
  Nucleosynthesis},'' \href{http://dx.doi.org/10.1103/PhysRevD.100.015038}{{\em
  Phys. Rev. D} {\bfseries 100} no.~1, (2019) 015038},
  \href{http://arxiv.org/abs/1904.04256}{{\ttfamily arXiv:1904.04256
  [hep-ph]}}.

\bibitem{CMB-S4:2016ple}
{\bfseries CMB-S4} Collaboration, K.~N. Abazajian {\em et~al.}, ``{CMB-S4
  Science Book, First Edition},''
  \href{http://arxiv.org/abs/1610.02743}{{\ttfamily arXiv:1610.02743
  [astro-ph.CO]}}.

\bibitem{Depta:2019lbe}
P.~F. Depta, M.~Hufnagel, K.~Schmidt-Hoberg, and S.~Wild, ``{BBN constraints on
  the annihilation of MeV-scale dark matter},''
  \href{http://dx.doi.org/10.1088/1475-7516/2019/04/029}{{\em JCAP} {\bfseries
  04} (2019) 029}, \href{http://arxiv.org/abs/1901.06944}{{\ttfamily
  arXiv:1901.06944 [hep-ph]}}.

\bibitem{Sabti:2019mhn}
N.~Sabti, J.~Alvey, M.~Escudero, M.~Fairbairn, and D.~Blas, ``{Refined Bounds
  on MeV-scale Thermal Dark Sectors from BBN and the CMB},''
  \href{http://dx.doi.org/10.1088/1475-7516/2020/01/004}{{\em JCAP} {\bfseries
  01} (2020) 004}, \href{http://arxiv.org/abs/1910.01649}{{\ttfamily
  arXiv:1910.01649 [hep-ph]}}.

\bibitem{Raffelt:1996wa}
G.~G. Raffelt, {\em {Stars as laboratories for fundamental physics}: {The
  astrophysics of neutrinos, axions, and other weakly interacting particles}}.
\newblock 5, 1996.

\bibitem{Grifols:1986fc}
J.~A. Grifols and E.~Masso, ``{Constraints on Finite Range Baryonic and
  Leptonic Forces From Stellar Evolution},''
  \href{http://dx.doi.org/10.1016/0370-2693(86)90509-5}{{\em Phys. Lett. B}
  {\bfseries 173} (1986) 237--240}.

\bibitem{1989MPLA....4..311G}
J.~A. {Grifols}, E.~{Mass{\'o}}, and S.~{Peris}, ``{Energy Loss from the Sun
  and Red Giants:. Bounds on Short-Range Baryonic and Leptonic Forces},''
  \href{http://dx.doi.org/10.1142/S0217732389000381}{{\em Modern Physics
  Letters A} {\bfseries 4} no.~4, (Jan., 1989) 311--323}.

\bibitem{DeRocco:2019jti}
W.~DeRocco, P.~W. Graham, D.~Kasen, G.~Marques-Tavares, and S.~Rajendran,
  ``{Supernova signals of light dark matter},''
  \href{http://dx.doi.org/10.1103/PhysRevD.100.075018}{{\em Phys. Rev. D}
  {\bfseries 100} no.~7, (2019) 075018},
  \href{http://arxiv.org/abs/1905.09284}{{\ttfamily arXiv:1905.09284
  [hep-ph]}}.

\bibitem{Tulin:2017ara}
S.~Tulin and H.-B. Yu, ``{Dark Matter Self-interactions and Small Scale
  Structure},'' \href{http://dx.doi.org/10.1016/j.physrep.2017.11.004}{{\em
  Phys. Rept.} {\bfseries 730} (2018) 1--57},
  \href{http://arxiv.org/abs/1705.02358}{{\ttfamily arXiv:1705.02358
  [hep-ph]}}.

\bibitem{Feng:2009hw}
J.~L. Feng, M.~Kaplinghat, and H.-B. Yu, ``{Halo Shape and Relic Density
  Exclusions of Sommerfeld-Enhanced Dark Matter Explanations of Cosmic Ray
  Excesses},'' \href{http://dx.doi.org/10.1103/PhysRevLett.104.151301}{{\em
  Phys. Rev. Lett.} {\bfseries 104} (2010) 151301},
  \href{http://arxiv.org/abs/0911.0422}{{\ttfamily arXiv:0911.0422 [hep-ph]}}.

\bibitem{Buckley:2009in}
M.~R. Buckley and P.~J. Fox, ``{Dark Matter Self-Interactions and Light Force
  Carriers},'' \href{http://dx.doi.org/10.1103/PhysRevD.81.083522}{{\em Phys.
  Rev. D} {\bfseries 81} (2010) 083522},
  \href{http://arxiv.org/abs/0911.3898}{{\ttfamily arXiv:0911.3898 [hep-ph]}}.

\bibitem{Loeb:2010gj}
A.~Loeb and N.~Weiner, ``{Cores in Dwarf Galaxies from Dark Matter with a
  Yukawa Potential},''
  \href{http://dx.doi.org/10.1103/PhysRevLett.106.171302}{{\em Phys. Rev.
  Lett.} {\bfseries 106} (2011) 171302},
  \href{http://arxiv.org/abs/1011.6374}{{\ttfamily arXiv:1011.6374
  [astro-ph.CO]}}.

\bibitem{Kaplinghat:2015aga}
M.~Kaplinghat, S.~Tulin, and H.-B. Yu, ``{Dark Matter Halos as Particle
  Colliders: Unified Solution to Small-Scale Structure Puzzles from Dwarfs to
  Clusters},'' \href{http://dx.doi.org/10.1103/PhysRevLett.116.041302}{{\em
  Phys. Rev. Lett.} {\bfseries 116} no.~4, (2016) 041302},
  \href{http://arxiv.org/abs/1508.03339}{{\ttfamily arXiv:1508.03339
  [astro-ph.CO]}}.

\bibitem{Agrawal:2020lea}
P.~Agrawal, A.~Parikh, and M.~Reece, ``{Systematizing the Effective Theory of
  Self-Interacting Dark Matter},''
  \href{http://dx.doi.org/10.1007/JHEP10(2020)191}{{\em JHEP} {\bfseries 10}
  (2020) 191}, \href{http://arxiv.org/abs/2003.00021}{{\ttfamily
  arXiv:2003.00021 [hep-ph]}}.

\bibitem{Parikh:2020ggm}
A.~Parikh, ``{The singularity structure of quantum-mechanical potentials},''
  \href{http://dx.doi.org/10.1103/physrevd.104.036005}{{\em Phys. Rev. D}
  {\bfseries 104} no.~3, (2021) 036005},
  \href{http://arxiv.org/abs/2012.11606}{{\ttfamily arXiv:2012.11606
  [hep-th]}}.

\bibitem{LZ:2022ufs}
{\bfseries LZ} Collaboration, J.~Aalbers {\em et~al.}, ``{First Dark Matter
  Search Results from the LUX-ZEPLIN (LZ) Experiment},''
  \href{http://arxiv.org/abs/2207.03764}{{\ttfamily arXiv:2207.03764
  [hep-ex]}}.

\bibitem{Cappiello:2018hsu}
C.~V. Cappiello, K.~C.~Y. Ng, and J.~F. Beacom, ``{Reverse Direct Detection:
  Cosmic Ray Scattering With Light Dark Matter},''
  \href{http://dx.doi.org/10.1103/PhysRevD.99.063004}{{\em Phys. Rev. D}
  {\bfseries 99} no.~6, (2019) 063004},
  \href{http://arxiv.org/abs/1810.07705}{{\ttfamily arXiv:1810.07705
  [hep-ph]}}.

\bibitem{Bringmann:2018cvk}
T.~Bringmann and M.~Pospelov, ``{Novel direct detection constraints on light
  dark matter},'' \href{http://dx.doi.org/10.1103/PhysRevLett.122.171801}{{\em
  Phys. Rev. Lett.} {\bfseries 122} no.~17, (2019) 171801},
  \href{http://arxiv.org/abs/1810.10543}{{\ttfamily arXiv:1810.10543
  [hep-ph]}}.

\bibitem{CRESST:2017ues}
{\bfseries CRESST} Collaboration, G.~Angloher {\em et~al.}, ``{Results on
  MeV-scale dark matter from a gram-scale cryogenic calorimeter operated above
  ground},'' \href{http://dx.doi.org/10.1140/epjc/s10052-017-5223-9}{{\em Eur.
  Phys. J. C} {\bfseries 77} no.~9, (2017) 637},
  \href{http://arxiv.org/abs/1707.06749}{{\ttfamily arXiv:1707.06749
  [astro-ph.CO]}}.

\bibitem{Ruppin:2014bra}
F.~Ruppin, J.~Billard, E.~Figueroa-Feliciano, and L.~Strigari,
  ``{Complementarity of dark matter detectors in light of the neutrino
  background},'' \href{http://dx.doi.org/10.1103/PhysRevD.90.083510}{{\em Phys.
  Rev. D} {\bfseries 90} no.~8, (2014) 083510},
  \href{http://arxiv.org/abs/1408.3581}{{\ttfamily arXiv:1408.3581 [hep-ph]}}.

\bibitem{Schutz:2016tid}
K.~Schutz and K.~M. Zurek, ``{Detectability of Light Dark Matter with
  Superfluid Helium},''
  \href{http://dx.doi.org/10.1103/PhysRevLett.117.121302}{{\em Phys. Rev.
  Lett.} {\bfseries 117} no.~12, (2016) 121302},
  \href{http://arxiv.org/abs/1604.08206}{{\ttfamily arXiv:1604.08206
  [hep-ph]}}.

\bibitem{Hertel:2018aal}
S.~A. Hertel, A.~Biekert, J.~Lin, V.~Velan, and D.~N. McKinsey, ``{Direct
  detection of sub-GeV dark matter using a superfluid $^4$He target},''
  \href{http://dx.doi.org/10.1103/PhysRevD.100.092007}{{\em Phys. Rev. D}
  {\bfseries 100} no.~9, (2019) 092007},
  \href{http://arxiv.org/abs/1810.06283}{{\ttfamily arXiv:1810.06283
  [physics.ins-det]}}.

\bibitem{Leane:2020wob}
R.~K. Leane and J.~Smirnov, ``{Exoplanets as Sub-GeV Dark Matter Detectors},''
  \href{http://dx.doi.org/10.1103/PhysRevLett.126.161101}{{\em Phys. Rev.
  Lett.} {\bfseries 126} no.~16, (2021) 161101},
  \href{http://arxiv.org/abs/2010.00015}{{\ttfamily arXiv:2010.00015
  [hep-ph]}}.

\bibitem{Leane:2022hkk}
R.~K. Leane and J.~Smirnov, ``{Floating Dark Matter in Celestial Bodies},''
  \href{http://arxiv.org/abs/2209.09834}{{\ttfamily arXiv:2209.09834
  [hep-ph]}}.

\bibitem{Gould:1988ym}
A.~P. Gould, ``{THE CAPTURE AND EVAPORATION OF WEAKLY INTERACTING MASSIVE
  PARTICLES},'' other thesis, 1988.

\bibitem{Gould:1988eq}
A.~Gould, J.~A. Frieman, and K.~Freese, ``{Probing the Earth With Wimps},''
  \href{http://dx.doi.org/10.1103/PhysRevD.39.1029}{{\em Phys. Rev. D}
  {\bfseries 39} (1989) 1029}.

\bibitem{Gould:1989hm}
A.~Gould and G.~Raffelt, ``{THERMAL CONDUCTION BY MASSIVE PARTICLES},''
  \href{http://dx.doi.org/10.1086/168568}{{\em Astrophys. J.} {\bfseries 352}
  (1990) 654}.

\bibitem{Acevedo:2023}
R.~K. Javier Acevedo~F., Leane and J.~Smirnov, ``{Evaporation Barrier for Dark
  Matter in Celestial Bodies},''
  \href{http://arxiv.org/abs/2201.1111X}{{\ttfamily arXiv:2201.1111X
  [hep-ph]}}.

\bibitem{Spergel:1984re}
D.~N. Spergel and W.~H. Press, ``{Effect of hypothetical, weakly interacting,
  massive particles on energy transport in the solar interior},''
  \href{http://dx.doi.org/10.1086/163336}{{\em Astrophys. J.} {\bfseries 294}
  (1985) 663--673}.

\bibitem{Knapen:2017xzo}
S.~Knapen, T.~Lin, and K.~M. Zurek, ``{Light Dark Matter: Models and
  Constraints},'' \href{http://dx.doi.org/10.1103/PhysRevD.96.115021}{{\em
  Phys. Rev. D} {\bfseries 96} no.~11, (2017) 115021},
  \href{http://arxiv.org/abs/1709.07882}{{\ttfamily arXiv:1709.07882
  [hep-ph]}}.

\bibitem{E949:2008btt}
{\bfseries E949} Collaboration, A.~V. Artamonov {\em et~al.}, ``{New
  measurement of the $K^{+} \to \pi^{+} \nu \bar{\nu}$ branching ratio},''
  \href{http://dx.doi.org/10.1103/PhysRevLett.101.191802}{{\em Phys. Rev.
  Lett.} {\bfseries 101} (2008) 191802},
  \href{http://arxiv.org/abs/0808.2459}{{\ttfamily arXiv:0808.2459 [hep-ex]}}.

\bibitem{MiniBooNE:2017nqe}
{\bfseries MiniBooNE} Collaboration, A.~A. Aguilar-Arevalo {\em et~al.},
  ``{Dark Matter Search in a Proton Beam Dump with MiniBooNE},''
  \href{http://dx.doi.org/10.1103/PhysRevLett.118.221803}{{\em Phys. Rev.
  Lett.} {\bfseries 118} no.~22, (2017) 221803},
  \href{http://arxiv.org/abs/1702.02688}{{\ttfamily arXiv:1702.02688
  [hep-ex]}}.

\bibitem{MiniBooNEDM:2018cxm}
{\bfseries MiniBooNE DM} Collaboration, A.~A. Aguilar-Arevalo {\em et~al.},
  ``{Dark Matter Search in Nucleon, Pion, and Electron Channels from a Proton
  Beam Dump with MiniBooNE},''
  \href{http://dx.doi.org/10.1103/PhysRevD.98.112004}{{\em Phys. Rev. D}
  {\bfseries 98} no.~11, (2018) 112004},
  \href{http://arxiv.org/abs/1807.06137}{{\ttfamily arXiv:1807.06137
  [hep-ex]}}.

\bibitem{Aalbers:2022dzr}
J.~Aalbers {\em et~al.}, ``{A Next-Generation Liquid Xenon Observatory for Dark
  Matter and Neutrino Physics},''
  \href{http://arxiv.org/abs/2203.02309}{{\ttfamily arXiv:2203.02309
  [physics.ins-det]}}.

\bibitem{Murata:2014nra}
J.~Murata and S.~Tanaka, ``{A review of short-range gravity experiments in the
  LHC era},'' \href{http://dx.doi.org/10.1088/0264-9381/32/3/033001}{{\em
  Class. Quant. Grav.} {\bfseries 32} no.~3, (2015) 033001},
  \href{http://arxiv.org/abs/1408.3588}{{\ttfamily arXiv:1408.3588 [hep-ex]}}.

\bibitem{Leeb:1992qf}
H.~Leeb and J.~Schmiedmayer, ``{Constraint on hypothetical light interacting
  bosons from low-energy neutron experiments},''
  \href{http://dx.doi.org/10.1103/PhysRevLett.68.1472}{{\em Phys. Rev. Lett.}
  {\bfseries 68} (1992) 1472--1475}.

\bibitem{Pokotilovski:2006up}
Y.~N. Pokotilovski, ``{Constraints on new interactions from neutron scattering
  experiments},'' \href{http://dx.doi.org/10.1134/S1063778806060020}{{\em Phys.
  Atom. Nucl.} {\bfseries 69} (2006) 924--931},
  \href{http://arxiv.org/abs/hep-ph/0601157}{{\ttfamily arXiv:hep-ph/0601157}}.

\bibitem{Nesvizhevsky:2007by}
V.~V. Nesvizhevsky, G.~Pignol, and K.~V. Protasov, ``{Neutron scattering and
  extra short range interactions},''
  \href{http://dx.doi.org/10.1103/PhysRevD.77.034020}{{\em Phys. Rev. D}
  {\bfseries 77} (2008) 034020},
  \href{http://arxiv.org/abs/0711.2298}{{\ttfamily arXiv:0711.2298 [hep-ph]}}.

\bibitem{Kamiya:2015eva}
Y.~Kamiya, K.~Itagami, M.~Tani, G.~N. Kim, and S.~Komamiya, ``{Constraints on
  New Gravitylike Forces in the Nanometer Range},''
  \href{http://dx.doi.org/10.1103/PhysRevLett.114.161101}{{\em Phys. Rev.
  Lett.} {\bfseries 114} (2015) 161101},
  \href{http://arxiv.org/abs/1504.02181}{{\ttfamily arXiv:1504.02181
  [hep-ex]}}.

\bibitem{PhysRevLett.103.153201}
N.~P. Mehta, S.~T. Rittenhouse, J.~P. D'Incao, J.~von Stecher, and C.~H.
  Greene, ``General theoretical description of $n$-body recombination,''
  \href{http://dx.doi.org/10.1103/PhysRevLett.103.153201}{{\em Phys. Rev.
  Lett.} {\bfseries 103} (Oct, 2009) 153201}.
  \url{https://link.aps.org/doi/10.1103/PhysRevLett.103.153201}.

\bibitem{XENON:2017vdw}
{\bfseries XENON} Collaboration, E.~Aprile {\em et~al.}, ``{First Dark Matter
  Search Results from the XENON1T Experiment},''
  \href{http://dx.doi.org/10.1103/PhysRevLett.119.181301}{{\em Phys. Rev.
  Lett.} {\bfseries 119} no.~18, (2017) 181301},
  \href{http://arxiv.org/abs/1705.06655}{{\ttfamily arXiv:1705.06655
  [astro-ph.CO]}}.

\bibitem{XENON:2018voc}
{\bfseries XENON} Collaboration, E.~Aprile {\em et~al.}, ``{Dark Matter Search
  Results from a One Ton-Year Exposure of XENON1T},''
  \href{http://dx.doi.org/10.1103/PhysRevLett.121.111302}{{\em Phys. Rev.
  Lett.} {\bfseries 121} no.~11, (2018) 111302},
  \href{http://arxiv.org/abs/1805.12562}{{\ttfamily arXiv:1805.12562
  [astro-ph.CO]}}.

\bibitem{Chatterjee:2019jts}
A.~Chatterjee, P.~Dayal, T.~R. Choudhury, and A.~Hutter, ``{Ruling out 3 keV
  warm dark matter using 21 cm EDGES data},''
  \href{http://dx.doi.org/10.1093/mnras/stz1444}{{\em Mon. Not. Roy. Astron.
  Soc.} {\bfseries 487} no.~3, (2019) 3560--3567},
  \href{http://arxiv.org/abs/1902.09562}{{\ttfamily arXiv:1902.09562
  [astro-ph.CO]}}.

\bibitem{KamLAND-Zen:2014hhc}
{\bfseries KamLAND-Zen} Collaboration, K.~Asakura {\em et~al.}, ``{Results from
  KamLAND-Zen},'' \href{http://dx.doi.org/10.1063/1.4915593}{{\em AIP Conf.
  Proc.} {\bfseries 1666} no.~1, (2015) 170003},
  \href{http://arxiv.org/abs/1409.0077}{{\ttfamily arXiv:1409.0077
  [physics.ins-det]}}.

\bibitem{JUNO:2015zny}
{\bfseries JUNO} Collaboration, F.~An {\em et~al.}, ``{Neutrino Physics with
  JUNO},'' \href{http://dx.doi.org/10.1088/0954-3899/43/3/030401}{{\em J. Phys.
  G} {\bfseries 43} no.~3, (2016) 030401},
  \href{http://arxiv.org/abs/1507.05613}{{\ttfamily arXiv:1507.05613
  [physics.ins-det]}}.

\bibitem{Hardy:2016kme}
E.~Hardy and R.~Lasenby, ``{Stellar cooling bounds on new light particles:
  plasma mixing effects},''
  \href{http://dx.doi.org/10.1007/JHEP02(2017)033}{{\em JHEP} {\bfseries 02}
  (2017) 033}, \href{http://arxiv.org/abs/1611.05852}{{\ttfamily
  arXiv:1611.05852 [hep-ph]}}.

\bibitem{Emken:2021lgc}
T.~Emken, ``{Solar reflection of light dark matter with heavy mediators},''
  \href{http://arxiv.org/abs/2102.12483}{{\ttfamily arXiv:2102.12483
  [hep-ph]}}.

\bibitem{Mei:2017etc}
D.~M. Mei, G.~J. Wang, H.~Mei, G.~Yang, J.~Liu, M.~Wagner, R.~Panth, K.~Kooi,
  Y.~Y. Yang, and W.~Z. Wei, ``{Direct Detection of MeV-Scale Dark Matter
  Utilizing Germanium Internal Amplification for the Charge Created by the
  Ionization of Impurities},''
  \href{http://dx.doi.org/10.1140/epjc/s10052-018-5653-z}{{\em Eur. Phys. J. C}
  {\bfseries 78} no.~3, (2018) 187},
  \href{http://arxiv.org/abs/1708.06594}{{\ttfamily arXiv:1708.06594
  [physics.ins-det]}}.

\bibitem{Hochberg:2017wce}
Y.~Hochberg, Y.~Kahn, M.~Lisanti, K.~M. Zurek, A.~G. Grushin, R.~Ilan, S.~M.
  Griffin, Z.-F. Liu, S.~F. Weber, and J.~B. Neaton, ``{Detection of sub-MeV
  Dark Matter with Three-Dimensional Dirac Materials},''
  \href{http://dx.doi.org/10.1103/PhysRevD.97.015004}{{\em Phys. Rev. D}
  {\bfseries 97} no.~1, (2018) 015004},
  \href{http://arxiv.org/abs/1708.08929}{{\ttfamily arXiv:1708.08929
  [hep-ph]}}.

\bibitem{Hochberg:2015fth}
Y.~Hochberg, M.~Pyle, Y.~Zhao, and K.~M. Zurek, ``{Detecting Superlight Dark
  Matter with Fermi-Degenerate Materials},''
  \href{http://dx.doi.org/10.1007/JHEP08(2016)057}{{\em JHEP} {\bfseries 08}
  (2016) 057}, \href{http://arxiv.org/abs/1512.04533}{{\ttfamily
  arXiv:1512.04533 [hep-ph]}}.

\bibitem{Marsicano:2018vin}
L.~Marsicano, M.~Battaglieri, A.~Celentano, R.~De~Vita, and Y.-M. Zhong,
  ``{Probing Leptophilic Dark Sectors at Electron Beam-Dump Facilities},''
  \href{http://dx.doi.org/10.1103/PhysRevD.98.115022}{{\em Phys. Rev. D}
  {\bfseries 98} no.~11, (2018) 115022},
  \href{http://arxiv.org/abs/1812.03829}{{\ttfamily arXiv:1812.03829
  [hep-ex]}}.

\bibitem{Batell:2014mga}
B.~Batell, R.~Essig, and Z.~Surujon, ``{Strong Constraints on Sub-GeV Dark
  Sectors from SLAC Beam Dump E137},''
  \href{http://dx.doi.org/10.1103/PhysRevLett.113.171802}{{\em Phys. Rev.
  Lett.} {\bfseries 113} no.~17, (2014) 171802},
  \href{http://arxiv.org/abs/1406.2698}{{\ttfamily arXiv:1406.2698 [hep-ph]}}.

\bibitem{Hanneke:2010au}
D.~Hanneke, S.~F. Hoogerheide, and G.~Gabrielse, ``{Cavity Control of a
  Single-Electron Quantum Cyclotron: Measuring the Electron Magnetic Moment},''
  \href{http://dx.doi.org/10.1103/PhysRevA.83.052122}{{\em Phys. Rev. A}
  {\bfseries 83} (2011) 052122},
  \href{http://arxiv.org/abs/1009.4831}{{\ttfamily arXiv:1009.4831
  [physics.atom-ph]}}.

\bibitem{Abi:2021gix}
{\bfseries Muon g-2} Collaboration, B.~Abi {\em et~al.}, ``{Measurement of the
  Positive Muon Anomalous Magnetic Moment to 0.46~ppm},''
  \href{http://dx.doi.org/10.1103/PhysRevLett.126.141801}{{\em Phys. Rev.
  Lett.} {\bfseries 126} no.~14, (2021) 141801},
  \href{http://arxiv.org/abs/2104.03281}{{\ttfamily arXiv:2104.03281
  [hep-ex]}}.

\bibitem{Emken:2019tni}
T.~Emken, R.~Essig, C.~Kouvaris, and M.~Sholapurkar, ``{Direct Detection of
  Strongly Interacting Sub-GeV Dark Matter via Electron Recoils},''
  \href{http://dx.doi.org/10.1088/1475-7516/2019/09/070}{{\em JCAP} {\bfseries
  09} (2019) 070}, \href{http://arxiv.org/abs/1905.06348}{{\ttfamily
  arXiv:1905.06348 [hep-ph]}}.

\bibitem{XENON:2019gfn}
{\bfseries XENON} Collaboration, E.~Aprile {\em et~al.}, ``{Light Dark Matter
  Search with Ionization Signals in XENON1T},''
  \href{http://dx.doi.org/10.1103/PhysRevLett.123.251801}{{\em Phys. Rev.
  Lett.} {\bfseries 123} no.~25, (2019) 251801},
  \href{http://arxiv.org/abs/1907.11485}{{\ttfamily arXiv:1907.11485
  [hep-ex]}}.

\bibitem{XENONCollaborationSS:2021sgk}
{\bfseries (XENON Collaboration)\textsection{}, XENON} Collaboration, E.~Aprile
  {\em et~al.}, ``{Emission of single and few electrons in XENON1T and limits
  on light dark matter},''
  \href{http://dx.doi.org/10.1103/PhysRevD.106.022001}{{\em Phys. Rev. D}
  {\bfseries 106} no.~2, (2022) 022001},
  \href{http://arxiv.org/abs/2112.12116}{{\ttfamily arXiv:2112.12116
  [hep-ex]}}.

\bibitem{XENONCollaboration:2022kmb}
{\bfseries (XENON Collaboration)\textdagger{}\textdagger{}, XENON}
  Collaboration, E.~Aprile {\em et~al.}, ``{Search for New Physics in
  Electronic Recoil Data from XENONnT},''
  \href{http://dx.doi.org/10.1103/PhysRevLett.129.161805}{{\em Phys. Rev.
  Lett.} {\bfseries 129} no.~16, (2022) 161805},
  \href{http://arxiv.org/abs/2207.11330}{{\ttfamily arXiv:2207.11330
  [hep-ex]}}.

\bibitem{An:2021qdl}
H.~An, H.~Nie, M.~Pospelov, J.~Pradler, and A.~Ritz, ``{Solar reflection of
  dark matter},'' \href{http://dx.doi.org/10.1103/PhysRevD.104.103026}{{\em
  Phys. Rev. D} {\bfseries 104} no.~10, (2021) 103026},
  \href{http://arxiv.org/abs/2108.10332}{{\ttfamily arXiv:2108.10332
  [hep-ph]}}.

\end{thebibliography}\endgroup

\end{document}